\documentclass[prd,twocolumn,showpacs,superscriptaddress,nofootinbib]{revtex4}

\usepackage{amsfonts}
\usepackage{amsmath}
\usepackage{amssymb}
\usepackage{upgreek}
\usepackage{bm}
\usepackage{dcolumn}
\usepackage{epsfig}
\usepackage{graphicx}
\usepackage{graphics}
\usepackage[latin1]{inputenc}
\usepackage{latexsym}
\usepackage{rotating}
\usepackage{hyperref}
\usepackage{xspace} 
\usepackage[usenames,dvipsnames]{color}

\usepackage{ulem}
\definecolor {darkgreen}{rgb}{0.2,0.7,0.2}

\newcommand\be{\begin{equation}}
\newcommand\ba{\begin{eqnarray}}
\newcommand\ee{\end{equation}}
\newcommand\ea{\end{eqnarray}}

\newcommand{\nn}{\nonumber}

\newcommand{\res}{{\mbox{\tiny Res}}}
\newcommand{\float}{{\mbox{\tiny float}}}
\newcommand{\final}{{\mbox{\tiny fin}}}
\newcommand{\initial}{{\mbox{\tiny ini}}}

\newcommand{\Newt}{{\mbox{\tiny Newt}}}
\newcommand{\PN}{{\mbox{\tiny PN}}}
\newcommand{\BD}{{\mbox{\tiny BD}}}
\newcommand{\Teuk}{{\mbox{\tiny Teuk}}}

\newcommand{\massless}{{\mbox{\tiny massless}}}
\newcommand{\massive}{{\mbox{\tiny massive}}}
\newcommand{\nonadiab}{{\mbox{\tiny non-adiab}}}
\newcommand{\Tot}{{\mbox{\tiny Total}}}
\newcommand{\ST}{{\mbox{\tiny ST}}}
\newcommand{\SCO}{{\mbox{\tiny SCO}}}
\newcommand{\MBH}{{\mbox{\tiny MBH}}}
\newcommand{\BH}{{\mbox{\tiny BH}}}
\newcommand{\ISCO}{{\mbox{\tiny ISCO}}}

\newcommand{\GW}{{\mbox{\tiny GW}}}

\newcommand{\GR}{{\mbox{\tiny GR}}}
\newcommand{\ppE}{{\mbox{\tiny ppE}}}

\newcommand{\TT}{{\mbox{\tiny TT}}}

\newcommand{\orb}{{\mbox{\tiny orb}}}
\newcommand{\obs}{{\mbox{\tiny obs}}}

\newcommand{\beq}{\begin{equation}}
\newcommand{\eeq}{\end{equation}}
\newcommand{\bes}{\begin{subequations}}
\newcommand{\ees}{\end{subequations}}
\newcommand{\beqn}{\begin{eqnarray*}}
\newcommand{\eeqn}{\end{eqnarray*}}

\newcommand{\f}[2]{\frac{#1}{#2}}

\def\nn{\nonumber}

\begin{document}
\title{Gravitational Waves from Quasicircular Extreme Mass-Ratio Inspirals \\ as Probes of Scalar-Tensor Theories} 
 
\author{Nicol\'as Yunes}
\affiliation{Department of Physics, Montana State University, Bozeman, MT 59717, USA.}

\author{Paolo Pani}
\affiliation{CENTRA, Departamento de F\'{\i}sica, 
Instituto Superior T\'ecnico, Universidade T\'ecnica de Lisboa - UTL,
Av.~Rovisco Pais 1, 1049 Lisboa, Portugal.}

\author{Vitor Cardoso}
\affiliation{CENTRA, Departamento de F\'{\i}sica, 
Instituto Superior T\'ecnico, Universidade T\'ecnica de Lisboa - UTL,
Av.~Rovisco Pais 1, 1049 Lisboa, Portugal.}
\affiliation{Department of Physics and Astronomy, The University of Mississippi, University, MS 38677, USA.}

\date{\today}

\begin{abstract} 

A stellar-mass compact object spiraling into a supermassive black hole, an extreme-mass-ratio inspiral, is one of the targets for future space-based gravitational-wave detectors. Such inspirals offer a unique opportunity to learn about astrophysics and test General Relativity in the strong-field. We here study whether scalar-tensor theories in asymptotically flat spacetimes can be further constrained with these inspirals. In the extreme-mass ratio limit, and assuming analyticity of the coupling functions entering the action, we show that all scalar-tensor theories universally reduce to massive or massless Brans-Dicke theory. We also show that in this limit, black holes do not emit dipolar radiation to all orders in post-Newtonian theory. For massless theories and quasi-circular orbits, we calculate the scalar energy flux in the test-particle, Teukolsky formalism to all orders in post-Newtonian theory and fit it to a high-order post-Newtonian expansion. We then derive the post-Newtonian corrections to the scalar-tensor modified, Fourier transform of the gravitational wave response function and map it to the parameterized post-Einsteinian framework. With the Teukolsky flux at hand, we use the effective-one-body framework adapted to extreme mass-ratio inspirals to calculate the scalar-tensor modifications to the gravitational waveform. We find that such corrections are smaller than those induced in the early inspiral of comparable-mass binaries, leading to projected bounds on the Brans-Dicke coupling parameter that are worse than current Solar System ones. This is because Brans-Dicke theory modifies the weak-field, leading to deviations in the energy flux that are largest at small velocities. For massive theories, superradiance can produce resonances in the scalar energy flux that can lead to quasi-circular floating orbits outside the innermost stable circular orbit and that last until the supermassive black hole loses enough mass and spin-angular momentum. If such floating orbits occur in the frequency band of a LISA-like mission, they would lead to a large dephasing (typically $\sim 10^6$ rads) that would prevent detection of such modified inspirals using General Relativity templates. A detection that is consistent with General Relativity would then rule out the presence of floating resonances at frequencies lower than the lowest inspiral frequency observed, allowing for the strongest constraints yet on massive scalar tensor theories. 

\end{abstract}




\maketitle

\section{Introduction}
Gravitational waves hold the promise to allow for new and stringent tests of General Relativity (GR) in a previously obscure regime: the strong-field region. This is the regime where gravitational fields are strong and the characteristic velocity of inspiraling bodies is a significant fraction of the speed of light. The inspiral, merger and ringdown of compact objects is the prototypical example of a gravitational wave source that emits in this region. We currently lack experimental verification of GR in the strong-field, as the most stringent tests we possess, thanks to the discovery of the double binary pulsar~\cite{2005ASPC..328...53B,Kramer:2006nb}, only probe the weak field. For example, the ratio of the total mass to the orbital separation (the gravitational compactness), which controls the magnitude of the orbital velocity, is of ${\cal{O}}(10^{-5}) \ll 1$ for the double binary pulsar~\cite{2005ASPC..328...53B,Kramer:2006nb}, while it is of ${\cal{O}}(1)$ during binary coalescence.

Much work has been devoted to studying whether gravitational waves from ground-based and space-based detectors could be used to test GR.  In most studies, one chooses a particular theory, or effect that one wishes to constrain, and then derives how gravitational waves would be deformed. Then, if a gravitational wave is detected that is consistent with GR, a constraint on gravitational wave deformations can be placed, which in turn are controlled by the magnitude of the coupling constants of the theory. Such a procedure has been applied to Brans-Dicke theory~\citep{Will:1994fb,Will:2004xi,Berti:2004bd,Berti:2005qd,Stavridis:2010zz,Arun:2009pq,Keppel:2010qu,Yagi:2009zm}, dynamical Chern-Simons gravity~\citep{Sopuerta:2009iy,Yunes:2009hc,Pani:2011xj,Yagi:2011xp}, modified quadratic gravity~\cite{Yagi:2011xp}, phenomenological massive graviton propagation~\citep{Will:1997bb,Scharre:2001hn,Will:2004xi,Berti:2004bd,Berti:2005qd,Yagi:2009zm,Berti:2011jz}, gravitational Lorentz-violation~\citep{Mirshekari:2011yq}, gravitational parity-violation~\citep{Yunes:2010yf,Yagi:2011xp}, violations of Local Position Invariance~\citep{Yunes:2009bv} and theories with extra-dimensions~\citep{Yagi:2011yu}. 

Most of these studies considered the early inspiral of compact objects within the post-Newtonian (PN) approximation, where fields are expanded in small velocities and weak-fields to high order in perturbation theory. Moreover, they typically considered only the leading-order (Newtonian) correction introduced by the given alternative theory and neglected all higher-order corrections. Such an approximation becomes particularly ill-suited in the late phase of inspiral, when the binary is close to the inner-most stable circular orbit (ISCO). This then suggests that highly relativistic systems might allow us to place different, perhaps more stringent constraints on modified gravity theories. 

An example of such a highly relativistic system is an extreme mass-ratio inspiral (EMRI), where a small compact object with mass $m_{\SCO} \in (1,10) M_{\odot}$ spirals into a supermassive black hole with mass $M_{\MBH} \in (10^{4},10^{7}) M_{\odot}$. The EMRI phase of interest to future space-based gravitational wave detection is the late inspiral, when the orbital separation is $\leq 20 M_{\MBH}$, the orbital period is $(10^{2},10^{4})$ seconds and low-frequency gravitational waves are emitted in the ($10^{-4},10^{-2}$) Hz frequency band. The merger and ringdown EMRI phases are completely negligible, as they contribute insignificant amounts of signal-to-noise ratio to the signal~\cite{Yunes:2009ke,2011PhRvD..84f2003C}. The EMRI inspiral alone can last tens to hundreds of years in the detector band, completely outlasting any realistic detection. In a single year of inspiral observation, millions of radians are contained in the signal, which encodes rich information about the spacetime dynamics~\cite{Ryan:1995wh}. EMRIs are therefore exceptional probes of strong-field GR and one might expect them to be the best probes of strong-field GR modifications. 

In this paper, we study whether the detection of equatorial and quasi-circular EMRIs allows for stronger constraints on a large class of scalar-tensor theories than those currently in place due to Solar System observations or those projected with future gravitational wave observations of comparable-mass binary inspirals. We choose to work with equatorial and quasi-circular EMRIs for two reasons. First, EMRI formation scenarios exist where the small compact object is in an equatorial and quasi-circular orbit by the time it enters the detector sensitivity band. One such model is that of~\cite{Levin:2006uc}, where a stellar-mass compact object is either created in an accretion disk surrounding a supermassive BH or is captured by the disk. The accretion disk is  expected to be in the spin equatorial plane inside a few hundred gravitational radii of the supermassive BH~\cite{Bardeen:1975zz}. Thus, in this scenario by the time the small compact object reaches sufficiently small separations to produce detectable gravitational waves, it would be in an equatorial and quasi-circular orbit. Second, although such EMRIs are less likely than inclined and eccentric ones (since other EMRI formation channels are more favourable to generic EMRIs), the tests we will discuss below only require a single detection. Nonetheless, the study presented here should lay the foundations for extensions to more generic orbits if needed.  

Scalar-tensor theories are prototypical models for non-minimally coupled scalar fields, which are common ingredients in low-energy effective actions of quantum gravity models (see e.g.~\cite{Fujii_Maeda_book}). Furthermore, scalar-tensor theories can be shown to include $f(R)$ gravity as a particular case (see e.g.~\cite{Sotiriou:2008rp,DeFelice:2010aj} for some reviews), thus being relevant as potential models for dark energy. Although here we focus on asymptotically flat geometries, our results should also apply to $f(R)$ theories, usually developed in asymptotically de-Sitter backgrounds, under the reasonable assumption that the cosmological background does not affect the local physics (see Ref.~\cite{Sotiriou:2011dz} for a discussion on this issue).

The most general, stationary, axisymmetric and vacuum solution to generic scalar-tensor theories is the Kerr metric coupled to a constant scalar field. In fact, Sotiriou and Faraoni~\cite{Sotiriou:2011dz} recently proved, under quite generic conditions, that such Kerr black holes are the endpoint of gravitational collapse in any scalar-tensor theory. This proof, however, fails for a cosmological, non-stationary (linear-in-time) scalar field~\cite{2011arXiv1111.4009H}, which might lead to violations of the no-hair theorem applied to scalar-tensor theories~\cite{2011arXiv1111.4009H,Bekenstein:1995un}. 

The Kerr metric will be adopted as the background spacetime upon which a small compact object inspirals. Although we mainly work in the Einstein frame, physical observables are obtained by mapping results to the physical Jordan frame. We show that, for a large class of scalar-tensor theories in an asymptotically flat, stationary background and expanding the fields about perturbations proportional to the mass ratio, the equations governing the emission of gravitational and scalar waves only depend on two parameters: a coupling constant $\alpha$ and the mass of the scalar field $\mu_s$, both of which are uniquely determined by the specific scalar-tensor model. Thus, one can think of scalar-tensor theories as a continuous (Lie) space (manifold) of theories, where $(\alpha,\mu_{s})$ are coordinates.

Brans-Dicke theory is contained in the subset of massless scalar-tensor theories, i.e.~the subset of scalar-tensor theories with $\mu_{s} = 0$. This theory not only requires a massless scalar field, but also that $\alpha$ be related to $\omega_{\BD}$ by a specific relation (cf.~Eq.~\eqref{def_alpha} below). Brans-Dicke theory is then parameterized by a single quantity, $\omega_{\BD}$, and can thus be thought of as a line in the $(\alpha,\mu_{s})$ space of all scalar-tensor theories. 

\section*{Executive Summary of Results}
\label{summary}

We study EMRIs in generic scalar tensor theories. By assuming analyticity of the coupling functions entering the action, we show that any of such scalar-tensor theories in the EMRI limit reduces identically and universally to massless or massive Brans-Dicke theory. To next-order in the mass ratio, this degeneracy in theory space breaks down. Therefore, any bound on Brans-Dicke theory derived from EMRI calculations automatically constrains the linear-in-mass-ratio expansion of all scalar-tensor theories. We also show that in the EMRI limit, black hole binaries in any scalar-tensor theory do not lead to dipolar radiation to all orders in PN theory, provided the background scalar field is constant at spatial infinity. 

We then study massless scalar-tensor theories. First, we numerically calculate the scalar energy flux carried to spatial infinity in the point-particle approximation within the Teukolsky formalism and assuming quasi-circular orbits. This calculation of the flux is fully relativistic, albeit valid only to leading order in the mass ratio, but accounting for all orders in PN theory. We fit this flux to a $3$PN and a $3.5$PN expansion and calculate the Fourier transform of the gravitational wave response function in the stationary-phase approximation beyond Newtonian order. We find that the PN corrections to the massless scalar-tensor modifications to the Fourier phase can be mapped to the recently proposed parameterized post-Einsteinian (ppE) framework~\cite{Yunes:2009ke,2011PhRvD..84f2003C} in a straightforward manner, which we explicitly provide. 

We then proceed to determine whether these PN corrections to the modified gravitational wave allow us to constrain the Brans-Dicke coupling parameter better than with comparable-mass binary inspirals. Recall that due to the equivalence discussed earlier, the bounds derived here also apply to all massless scalar-tensor theories to leading-order in the mass ratio. With the Teukolsky flux at hand, we employ the effective-one-body (EOB) framework~\cite{Buonanno99,Buonanno00}, recently adapted to EMRIs~\cite{2009GWN.....2....3Y,Yunes:2010zj}, to evolve inspiral with and without the Brans-Dicke correction. We then compare their associated gravitational wave phases, after minimizing the difference over an arbitrary time and phase offset. Contrary to our expectation, we find that EMRIs will not be able to constrain Brans-Dicke theory beyond current Solar System constraints. In fact, the projected bounds we obtain are {\emph{worse}} than those found by considering gravitational-wave detection of comparable-mass binary inspirals with only the leading-order Brans-Dicke correction~\citep{Will:1994fb,Will:2004xi,Berti:2005qd,Stavridis:2010zz,Arun:2009pq,Keppel:2010qu,Yagi:2009zm}. 

This seemingly surprising result is in fact perfectly reasonable, once one realizes that scalar-tensor theories are not necessarily strong-curvature modifications of GR. At the level of the action, the scalar-field modification acts on the Ricci scalar in the Jordan frame and it does not introduce higher curvature corrections. This is consistent with the fact that scalar-tensor theories include $f(R)$ gravity, which is usually studied as an infrared, rather than an ultraviolet (or strong-field/strong-curvature) correction to GR. In terms of the energy flux, the weak-field nature of massless scalar-tensor theories is evidenced by the scalar flux becoming less and less important than the GR one as the small compact object approaches the horizon. This is why the modified theory introduces a {\emph{pre}}-Newtonian correction to the gravitational wave phase (a $-1$PN effect relative to GR), instead of a {\emph{post}}-Newtonian one.

\begin{figure}
\epsfig{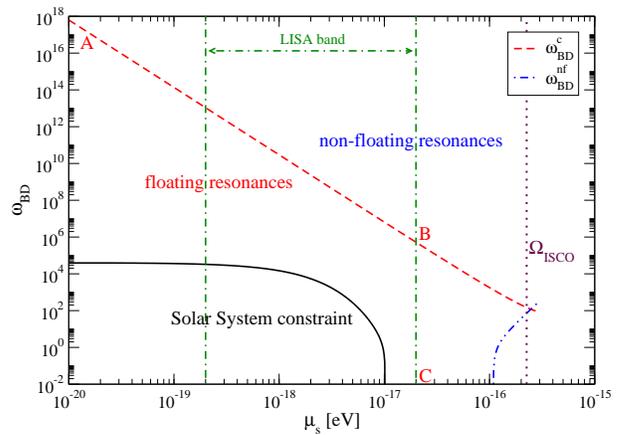}
 \caption{\label{fig:floating-regions} 
Parameter space of massive Brans-Dicke theory for a quasi-circular EMRI with typical neutron star sensitivity $s_\SCO=0.188$, supermassive black hole spin $a_\MBH/M_{\MBH}=0.9$ and mass $M_\MBH=10^5 M_\odot$. The red, dashed curve is the boundary of the region in $(\omega_\BD,\mu_{s})$ space that separates floating and non-floating resonances (cf. Eq.~\eqref{omegaBD_crit_Einstein}). The area below the black curve shows the region that is ruled out by Solar System experiments~\cite{Perivolaropoulos:2009ak,Alsing:2011er}. The area between the green, dot-dashed curves shows the region where a resonance would occur inside the classic LISA sensitivity band [$(10^{-4},10^{-2})$ Hz]. Due to a large dephasing introduced by floating resonances, any detection of an EMRI in a LISA-like mission would rule out the (open) region delimited by ABC. The dot-dashed blue line shows the values of $(\omega_\BD,\mu_{s})$ below which non-floating resonances lead to dephasings larger than $1$ rad (see Sec.~\ref{sec:resonant-effect-on-GW}). The ISCO frequency, $\Omega_\ISCO$, is shown by a vertical dotted line. The curves $\omega_\BD^{c}$ and $\omega_{\BD}^{\rm nf}$ terminate at $\mu_s\sim\Omega_\orb\sim\Omega_\ISCO\sim0.22/M_\MBH$. }
\end{figure}

With the massless case under control, we then proceed to consider massive scalar-tensor theories, where now GR deformations depend both on the mass of the scalar field $\mu_{s}$ and the scalar-tensor coupling parameter $\alpha$. Depending on the values of $(\alpha,\mu_{s})$, prograde EMRI orbits can experience superradiance and produce resonances in the scalar flux that can momentarily counteract the gravitational wave flux, leading to {\emph{floating orbits}}~\cite{Cardoso:2011xi}, i.e.~orbits where the inspiral greatly slows down at a given floating radius. Figure~\ref{fig:floating-regions} shows the separatrix (red dashed line) between floating and non-floating orbits for an EMRI with total mass $M\sim10^{5}M_{\odot}$ in massive Brans-Dicke theory. As indicated in this figure, floating orbits can exist only in the region above and to the right of the red-dashed line. This curve (and all others) terminate when the small compact object approaches the ISCO, approximately when $\mu_s\sim\Omega_\ISCO$, where $\Omega_\ISCO$ is the ISCO frequency (vertical dotted line in Fig.~\ref{fig:floating-regions}).

We have chosen to show the region of allowed floating orbits in $(\omega_{\BD},\mu_{s})$ coordinates instead of $(\alpha,\mu_{s})$, thus restricting attention to massive Brans-Dicke theory via Eq.~\eqref{def_alpha},  so as to be able to compare directly with Solar System experiments~\cite{Bertotti:2003rm,Perivolaropoulos:2009ak}. The area under the solid black line in Fig.~\ref{fig:floating-regions} denotes the region that is currently ruled out by such experiments~\cite{Bertotti:2003rm,Perivolaropoulos:2009ak}. Observe that as $\mu_{s} \to 0$, the Solar System constraint becomes independent of $\mu_s$ and approaches $\omega_{\BD} \geq 4 \times 10^{4}$, as expected. For completeness, we also show the region (between green, dot-dashed vertical lines) for which the floating orbits would lead to GWs inside the classic-LISA sensitivity band [$(10^{-4},10^{-2})$ Hz], since the resonant frequency is linearly proportional to $\mu_{s}$(see Ref.~\cite{AmaroSeoane:2012km} for a recent review on eLISA-NGO). From this, one can easily observe that there is a large region in $(\omega_{\BD},\mu_{s})$ space that is not ruled out by Solar System experiments, yet where floating orbits should be observed with LISA like instruments, ie.~the region between the greened dot-dashed curves, above the black solid line and below the red dashed line. As we explain below, the modifications to an EMRI GW signal due to hitting a floating orbit are so large that an observation of a signal consistent with GR would rule out the region below the red dashed line, thus improving Solar System constraints by several orders of magnitude.  

Floating orbits are not a new invention in the context of EMRIs and GR. Press and Teukolsky~\cite{1972Natur.238..211P,Thorne:1986iy}, after a suggestion of Misner's, were the first to study floating orbits within GR. The main idea is that for certain orbital configurations, a small object orbiting a supermassive black hole can {\emph{extract}} angular momentum from the latter. That is, the gravitational waves that impinge on the supermassive black hole horizon are not absorbed, but instead reflected with greater energy and angular momentum that they possessed initially. If such extraction is sufficiently large to balance the energy lost by the system due to gravitational waves traveling out to spatial infinity, the small object might momentarily experience no radiation-reaction force and stall in its orbit.  Press and Teukolsky found that, although such floating orbits are not impossible in GR, for them to exist outside the ISCO of a prograde orbit, one requires an essentially maximally spinning supermassive black hole, with spins exceeding the Thorne limit $|\vec{S}_{\rm max}| = 0.998 M_{\MBH}^{2}$~\cite{1972Natur.238..211P,Thorne:1986iy,1974ApJ...191..507T}. In massive scalar-tensor theory, more energy can be extracted due to the activation of the scalar energy flux, thus allowing for floating orbits outside the ISCO for supermassive black holes with smaller spins. 

Floating orbits, however, are somewhat of a misnomer, as an EMRI that attempts to cross such a floating radius does not completely stall or stagnate as the name implies. Instead, as an EMRI attempts to cross a floating radius, its evolution is greatly slowed down, as it is now driven by the supermassive black hole's energy and spin angular momentum loss. In other words, the radiation-reaction force or the energy flux are weakened by an additional factor of mass-ratio $\nu$. Thus, while to leading (first) order the energy flux scales with $\nu^{2}$, during floating it scales with $\nu^{3}$, i.e.~it becomes a second-order in the mass ratio effect.

The time it takes the EMRI to cross a floating resonance is determined by the supermassive black hole's shedding efficiency [of ${\cal{O}}(10^{6})$ yrs for a typical EMRI]. This timescale can easily exceed a Hubble time for EMRIs orbiting at separations greater than $326 M_{\MBH}$. While the EMRI is slowed down, it will produce gravitational waves very different from those in GR, leading to a large dephasing, i.e.~a phase difference relative to the GR phase, of roughly $10^6$ rads over a one year observation for phenomenologically viable $(\alpha,\mu_{s})$ parameters. Therefore, if massive, scalar-tensor theories are the correct representation of Nature, such a large dephasing would prevent detection of EMRIs in a GR template-based, matched-filtering gravitational wave search.   

On the other hand, the detection of an EMRI gravitational wave that is consistent with GR can then be used to constrain massive scalar-tensor theories. This is because such a detection would imply that a resonance was not present in the range of frequencies sampled by the EMRI. If a resonance were present at a frequency lower than the lowest one detected, the EMRI would have hanged at the resonance frequency for a very long time, evolving very slowly, and never entering the detector's sensitivity band. The requirement that the resonant frequency be lower than the lowest EMRI frequency detected would impose unprecedented constraints on massive scalar tensor theory: the (open) region delimited by ABC in Fig.~\ref{fig:floating-regions} would be ruled out. Observe that such a constraint is much stronger than current Solar-System ones (solid black line in Fig.~\ref{fig:floating-regions}). Therefore, even though future space-based gravitational wave detectors will not be able to constrain massless scalar-tensor theories beyond current Solar System levels~\cite{Bertotti:2003rm,Perivolaropoulos:2009ak}, they will be able to severely bound a large sector of massive scalar tensor theories that is today still phenomenologically viable.

The remainder of this paper presents the details of this calculation. 
Section~\ref{sec:review} reviews the basics of scalar-tensor theories applicable to EMRIs.
Section~\ref{sec:GWST} describes the test-particle approximation and the Teukolsky framework used to model EMRIs in scalar-tensor theories.
Section~\ref{Sec:EfluxMassless} calculates the energy flux in the Teukolsky formalism for massless scalar fields, fits the massless flux to a PN expansion and computes the Fourier transform of the response function in the stationary-phase approximation. 
Section~\ref{Sec:EfluxMassive} repeats the analysis of Sec.~\ref{Sec:EfluxMassless} but for massive scalar fields.
Section~\ref{sec:Constraints} describes how to model EMRIs using the EOB-EMRI framework both in GR and in massless scalar-tensor theories, and then calculates the dephasing between a GR waveform and a massless scalar-tensor theory one, allowing us to derive a projected bound. 
Section~\ref{sec:Constraints-Massive} derives analytical estimates on how well massive scalar-tensor theories will be constrained with future gravitational wave detectors. 
Section~\ref{sec:conclusions} concludes and points to future research. 

Throughout the rest of this paper we follow mainly the conventions of Misner, Thorne and Wheeler~\cite{Misner:1973cw}. We use Greek letters to denote spacetime indices, Latin ones at the middle of the alphabet $i,j,\ldots$ stand for spatial indices only, and Latin ones at the beginning of the alphabet $a,b,\ldots$ range over waveform parameters. We also use geometric units with $G=c=\hbar=1$, except in Section~\ref{sec:review} where we restore $G$ for clarity. Here $G$ is the gravitational constant as measured by Cavendish-like experiments, i.e. as measured by a distant observer in the Jordan frame. 

\section{Scalar-Tensor Theories for Extreme Mass-Ratio Inspirals}
\label{sec:review}

In this section, we review the basics of scalar-tensor theories as they concern this paper. For a detailed review refer to~\cite{Fujii_Maeda_book,lrr-2006-3} and references therein. Consider then the action of a generic scalar tensor theory in the {\emph{Jordan frame}}:
\ba
S_{(J)}&&=\frac{1}{16\pi}\int d^4x \sqrt{-g}\left[F(\phi)R-Z(\phi)g^{\mu\nu}\partial_{\mu}\phi\partial_{\nu}\phi\right.\nn\\
&&\left.-2U(\phi)\right]+S_{\rm mat}(\Psi_m;g_{\mu\nu})\,,
\label{action_Jordan}
\ea
where $F(\phi)$, $Z(\phi)$ and $U(\phi)$ are functionals of the scalar field $\phi$ that modify the Einstein-Hilbert term, the minimal kinetic term and the scalar field potential. The quantity $S_{\rm mat}(\Psi_{m};g_{\mu \nu})$ denotes the action for other matter degrees of freedom. As usual, $R$ is the Ricci scalar and $g$ is the determinant of the physical metric $g_{\mu \nu}$. 

In the following, we employ a test-particle approximation, i.e.~we assume the symmetric mass ratio $\nu \equiv m_{\SCO} M_{\MBH}/(m_{\SCO}+M_{\MBH})^{2} \ll 1$, where $m_{\SCO}$ is the physical mass of a small compact object, while $M_{\MBH}$ is the physical mass of the supermassive (background) black hole. Thus, we consider the matter action for a point-particle moving on a fixed background
\begin{equation}
 S_{\rm mat}=-\int d\tau \; m(\phi)\,,
\end{equation}
where $\tau$ is the proper time along the trajectory. Scalar-tensor theories in the Jordan frame promote the mass of small object to a $\phi$-dependent quantity $m=m(\phi)$~\cite{1971ApJ...163..611W,1975ApJ...196L..59E}. This allows for possible effects on the structure of the compact object, and on its motion, due to the scalar field~\cite{Damour:1992we,lrr-2006-3,Ohashi:1996uz}. 

Let us now transform this action to the Einstein frame through the following conformal transformation~\cite{Damour:1996ke}
\ba
 g^{(E)}_{\mu\nu}&=&F(\phi)g_{\mu\nu}\,,\label{confg}\\
 \Phi(\phi)&=&\frac{1}{\sqrt{4\pi}}\int d\phi\,\left[\frac{3}{4}\frac{F'(\phi)^2}{F(\phi)^2}+\frac{1}{2}\frac{Z(\phi)}{F(\phi)}\right]^{1/2}\,,\label{confPhi}\\
 A(\Phi)&=&F^{-1/2}(\phi)\,,\label{confA}\\
 V(\Phi)&=&\frac{2 U(\phi)}{F^2(\phi)}\,.\label{confV}
\ea
The action in Einstein frame then becomes
\ba
 S_{(E)}&&=\int d^4x \sqrt{-g^{(E)}}\left(\frac{R^{(E)}}{16\pi}-\frac{1}{2}g^{(E)}_{\mu\nu}\partial^\mu\Phi\partial^\nu\Phi-\frac{V(\Phi)}{16\pi}\right)\nn\\
&&-\int d \tau^{(E)} A(\Phi)m(\phi)\,,\label{actionEinstein}
\ea
where we have used $d\tau=A(\Phi)d\tau^{(E)}$. The associated modified field equations are~\cite{Ohashi:1996uz}
\begin{align}
 G_{\mu\nu}^{(E)}&=8\pi \left(T_{\mu\nu}^{(E)}+ T_{\mu \nu}^{(\Phi)} \right)\,,
\\
\square^{(E)}{\Phi} &= \frac{1}{16 \pi} \frac{\partial V}{\partial \Phi}+\frac{1}{\sqrt{-g^{(E)}}}
\nn\\
&\times\int d\tau^{(E)}\frac{\partial \left[A(\Phi)m(\phi)\right]}{\partial \Phi} \delta^{(4)}\left[x^{\mu}-y^{\mu}(\tau^{(E)})\right]\,,\nn
\end{align}
where $y^{\mu}$ denotes the location of the test-particle, ${T^{\mu\nu}}^{(E)}={2}(-g^{(E)})^{-1/2}\delta S_{\rm mat}/\delta g^{(E)}_{\mu\nu}$ is the stress-energy tensor associated with the matter action, and  
\be
T_{\mu\nu}^{(\Phi)} = \partial_\mu\Phi\partial_\nu\Phi-\frac{1}{2} g_{\mu\nu}^{(E)} (\partial\Phi)^2 -\frac{1}{16 \pi} g_{\mu\nu}^{(E)} V(\Phi)\,.\nn\\
\ee
is the scalar field energy-momentum tensor. Notice that the field equations depend only on three generic functions, $V(\Phi)$, $A(\Phi)$ and $m(\phi)$. 

Let us now consider scalar perturbations around a constant background scalar field, $\Phi^{(0)}$, and around a metric which is solution of Einstein's equations in an asymptotically-flat spacetime. Hence, we expand about $\Phi-\Phi^{(0)}=\varphi\ll1$ \emph{assuming}\footnote{By assuming $V(\Phi)$, $A(\Phi)$ and $m(\Phi)$ to be analytic functions of $\Phi$ we exclude some viable $f(R)$ cosmological models which are equivalent to scalar-tensor theories with a divergent potential, $V(\Phi\to\Phi^{(0)})\to\infty$ (see e.g. Ref.~\cite{DeFelice:2010aj}). However, in the context of strong-field corrections to GR, the analyticity assumption is general enough to encompass a large (in fact, infinitely dimensional) class of viable theories in asymptotically flat spacetime.} a general analytical behavior of the potentials around $\Phi\sim \Phi^{(0)}$:
\ba
 V(\Phi)=\sum_{n=0}V_n\left(\Phi-\Phi^{(0)}\right)^n\,,\\
 A(\Phi)=\sum_{n=0}A_n\left(\Phi-\Phi^{(0)}\right)^n\,,\\
 m(\Phi)=\sum_{n=0}m_n\left(\Phi-\Phi^{(0)}\right)^n\,,
\ea
where $(V_{n},A_{n},m_{n})$ are constant coefficients independent of $\Phi$ and $m=m[\phi(\Phi)]=m(\Phi)$. 
Expanding the modified field equations in this way, we find
\ba
G_{\mu \nu}^{1,(E)} &=& 8\pi T_{\mu\nu}^{1,(E)}-\frac{1}{2}g_{\mu\nu}^{(E)}\left[V_0+V_1\varphi\right]\,,\\
\left[\square^{1,(E)}-\frac{V_2}{8\pi}\right] \varphi &=& \frac{V_1}{16\pi}+\left[\frac{A_{1}}{A_{0}} + \frac{m_1}{m_0}\right]T^{1,(E)}\,.
\ea
where
\begin{equation}
 T^{1,(E)}=A_0m_0\int \frac{d\tau^{(E)}}{\sqrt{-g^{0,(E)}}} \delta^{(4)}\left[x^{\mu}-y^{\mu}\left(\tau^{(E)}\right)\right]\,,\label{traceT_E}
\end{equation}
is the trace of the stress-energy tensor for a test particle in the Einstein frame. We have here dropped terms of ${\cal{O}}(\varphi \; \nu)$ and the sub or superscripts $0$ or $1$ refer to quantities that are truncated at zero or first order, respectively. Asymptotically flat backgrounds require $V_0=0=V_1$, or equivalently $U_{0} = 0 = U_{1}$.

Our final equations then read
\ba
  G_{\mu\nu}^{1,(E)}&=&{8\pi} T_{\mu\nu}^{1,(E)}\,,\label{EQBD1a}\\
 \left[\square^{1,(E)}-\mu_s^2\right]\varphi&=& \alpha \; T^{1,(E)}\label{EQBD2a}\,.
\ea
where we have defined
\be
 \alpha \equiv \frac{A_{1}}{A_{0}}+ \frac{m_1}{m_0}\,, 
 \qquad \mu_s^2 \equiv\frac{V_2}{8\pi}\,,\label{def_alpha_ST}
\ee
which are both constant. Notice that this is nothing but the Einstein equations for a particle with renormalized mass $m_{\SCO} = A_{0} m_{0}$. Thus, any measurement of the mass of the particle is in reality a measurement of the renormalized mass $m_{\SCO}$, instead of the bare mass $m_{0}$ in the Jordan frame. 

Given this, the gravitational constant as measured by Cavendish type experiments (i.e. by observers at infinity in the Jordan frame) reads~\cite{Damour:1992we}
\begin{equation}
 G=\left[1+\frac{1}{4\pi}\left(\frac{A'(\Phi)}{A(\Phi)}\right)^2\right]A(\Phi)^2\sim A_0^2+\frac{A_1^2}{4\pi}\,, \label{G_ST}
\end{equation}
whereas the ``sensitivity'' of the compact object is~\cite{1989ApJ...346..366W}
\begin{eqnarray}
 s_\SCO&\equiv&-\frac{d \log m}{d\log G}=-\frac{G(\Phi)}{m(\Phi)}\frac{m'(\Phi)}{G'(\Phi)}\nn\\
&\sim& -\frac{\pi m_{1} G}{A_{1} m_{0}} \left(A_2 +2 \pi A_0 \right)^{-1}\,. \label{sensitivity_ST}
\end{eqnarray}
In the next sections, following the convention of previous works~\cite{1989ApJ...346..366W,Ohashi:1996uz}, we shall use units such that $G=1$ which, by Eq.~\eqref{G_ST}, implies $A_0^2=1-A_1^2/(4\pi)$. This then allows us to simplify the expressions for the sensitivities in Eq.~\eqref{sensitivity_ST} and the $\alpha$ parameter in Eq.~\eqref{def_alpha_ST}. 

If the small compact object is a black hole, then its mass $m_\BH^{(E)}$ in the Einstein frame is constant~\cite{Jacobson:1999vr}. Using the mapping to the Jordan frame, $m_\BH=m^{(E)}_\BH/A(\Phi)$, one then finds 
\begin{equation}
 m_\BH\sim \frac{m_\BH^{(E)}}{A_0^2}(A_0-A_1\varphi) \quad \Rightarrow \quad \frac{m_1}{m_0}=-\frac{A_1}{A_0}\,, \label{massBH}
\end{equation}
and the black hole sensitivity reads
\begin{equation}
s_\BH=\frac{1}{2}\left(A_0^2+\frac{A_2 A_0}{2\pi}\right)^{-1}\,.\label{BHsensitivity}
\end{equation}
Although $s_\BH$ generically depends on the specific scalar-tensor theory and can be different from its value in Brans-Dicke theory~\cite{1989ApJ...346..366W} ($s_\BH=1/2$, as discussed below) if the small compact object is a black hole we get, by Eqs.~\eqref{def_alpha_ST} and \eqref{massBH}, $\alpha=0$. Thus, for \emph{any} (analytic) scalar-tensor theory, scalar perturbations \emph{decouple} from the matter sector and the energy and angular momentum emission is simply governed by Eq.~\eqref{EQBD1a}, which is the same as in GR. Notice that this result is valid to all orders in PN theory, but only to leading order in the mass ratio. However, this result has been recently confirmed for equal-mass binary black hole mergers with full numerical relativity simulations~\cite{Healy:2011ef}.  

After choosing $G=1$, the gravitational and scalar wave emission in the test-particle approximation only depends on two parameters, $\alpha$ and $\mu_s$, which are fixed by the specific scalar-tensor theory under consideration. Physically, $\alpha$ is related to the coupling function between the scalar field and the gravity sector, whereas $\mu_{s}$ is the mass of the scalar field. The latter is related to the parameter of the coupling functions in the Jordan frame via
\begin{equation}
 \mu_s=\frac{A_0^2 \phi'(\Phi^{(0)})}{2\sqrt{2\pi}}m_s\,,\label{mass_Jordan}
\end{equation}
where $m_s$ is defined by $U(\phi)=m_s^2(\phi-\phi_0)^2/2$. Note that, due to the non-minimal kinetic term $Z(\phi)$, the physical mass of the scalar field is $\mu_s$, and not $m_s$ (see e.g. Ref.~\cite{Alsing:2011er} for a related discussion in massive Brans-Dicke theory).

The modified field equations of scalar-tensor theories in the Einstein frame~\eqref{EQBD1a}-\eqref{EQBD2a} are equivalent to those arising in GR coupled to a free scalar field if one expands in scalar perturbations and in the mass ratio. Thus, the gravitational metric perturbation only has the usual two propagating degrees of freedom. This is in contrast to modified field equations in Jordan frame, where the metric perturbation has three degrees of freedom: the two usual Einstein ones plus a breathing mode~\cite{1989ApJ...346..366W}. 

In the next sections, we will study EMRIs directly in the Einstein frame and, once a solution is found, we will map the observables back to the Jordan frame. In the latter, the gravitational field at infinity reads
\begin{eqnarray}
 g_{\mu\nu}&=&A(\Phi)^2 g_{\mu\nu}^{(E)}=A(\Phi)^2 (\eta_{\mu\nu}^{(E)}+h_{\mu\nu}^{(E)})\nn\\
&\sim& A_0^2(\eta_{\mu\nu}^{(E)}+h_{\mu\nu}^{(E)}) + 2 A_1 A_0\varphi \eta_{\mu\nu}^{(E)}\nn\\
&\sim&\eta_{\mu\nu}+h_{\mu\nu}+ 2 \frac{A_1}{A_0} \varphi \eta_{\mu\nu} \,,\label{gmunuJordan}
\end{eqnarray}
where we have neglected terms of ${\cal O}(\varphi\, \nu)$ and quantities without superscript are in the Jordan frame. When computing waveforms, we are interested in the transverse-traceless part of $g_{\mu\nu}$ only, hence the only relevant term in Eq.~\eqref{gmunuJordan} is $h_{\mu\nu}^{\TT}=A_0^2 h_{\mu\nu}^{\TT,(E)}$. 

The quantity $A_{0}$ is a constant that modifies the amplitude of the transverse-traceless, metric perturbation, and thus, the gravitational fluxes of energy and angular momentum and the gravitational wave phase evolution. As discussed below, in massless Brans-Dicke theory, this is very close to unity: $A_{0}= [(2 \omega_{\BD} + 3 )/(2 \omega_{\BD} + 4)]^{1/2} \sim 1 - (4 \omega_{\BD})^{-1}$, where $\omega_{\BD}$ is the Brans-Dicke coupling parameter and in the last line we expanded about $\omega_{\BD} = \infty$ since $\omega_{\BD} \geq 4 \times 10^{4}$ to agree with Solar System experiments. Even in massive Brans-Dicke theory, where $\omega_{\BD}$ can evade Solar system constraints for a sufficiently large mass $\mu_{s}$, $A_{0}$ is still close to unity: if $\omega_{\BD} = 1$,  then $A_{0} = \sqrt{5}/3 \sim 0.745$. 

Regardless of its magnitude, the quantity $A_{0}$ has the effect of renormalizing the overall amplitude of the gravitational metric perturbation. This occurs because in Eq.~\eqref{G_ST} we have implicitly chosen $G^{(E)}=1$ in the Einstein frame. We could have, however, chosen $G^{(E)}$ in such a way as to force $A_{0} = 1$ exactly. In any case, for an EMRI observation, such an overall $(G^{(E)} A_{0})^{2}$ factor cannot be separately measured; it is degenerate with, for example, the luminosity distance. For this reason, we will ignore this overall factor of $A_{0}$ in remaining sections of this paper.

\subsubsection*{Massless Brans-Dicke Limit}

Brans-Dicke theory can be recovered from the field equations~\eqref{EQBD1a}-\eqref{EQBD2a} (i.e.~from the field equations in the test-particle approximation) in the massless, $\mu_s=0$, case with the choices
\ba
F(\phi)=\phi\,,\qquad Z(\phi)= \frac{\omega_{\BD}}{\phi}\,, \qquad U(\phi)=0\,, 
\label{masslessBD}
\ea
where $\omega_\BD$ is a constant.
This then leads to the conformal transformation~\cite{Ohashi:1996uz}
\ba
g_{\mu\nu}^{(E)} &=& \phi \; g_{\mu\nu}\,, 
\qquad
\Phi = \frac{1}{\beta}\ln \phi\,,
\nn \\
A(\Phi) &=& e^{-\beta\Phi/2}\,,
\qquad
T_{\mu\nu}^{(E)} = \frac{1}{\phi} T_{\mu\nu}\,,
\ea
where $\beta=\sqrt{16\pi/(2\omega_{\BD}+3)}$. In this case, the perturbation quantities $A_{i}$ read
\ba
A_{0} &=& \frac{1}{\sqrt{\phi_{0}}}\,,
\quad
A_{1} = -\frac{\beta}{2 \sqrt{\phi_{0}}}\,
\quad 
A_{2} = \frac{\beta^{2}}{8 \sqrt{\phi_{0}}}\,,
\ea
and Eq.~\eqref{G_ST} reduces to
\begin{equation}
 G=\frac{1}{\phi_0} \frac{2\omega_\BD+4}{2\omega_\BD+3}\,.
\end{equation}
Note that, if $G=1$ units are used, the equation above defines $\phi_0$ and $A_0$ in terms of $\omega_\BD$:
\be
\phi_{0} = \frac{2\omega_\BD+4}{2\omega_\BD+3}=A_0^{-2}\,.
\label{def_A0}
\ee

Finally, the field equations in Einstein frame for an asymptotically-flat spacetime are given by Eqs.~\eqref{EQBD1a} and~\eqref{EQBD2a} with
\be
\alpha = \sqrt{\frac{16 \pi G}{2\omega_\BD+3}} \left(s_{\SCO}-\frac{1}{2}\right)\,,
\qquad
\mu_s =0\,.
\label{def_alpha}
\ee
where the sensitivity of Eq.~\eqref{sensitivity_ST} reduces to 
\begin{equation}
 s_{\SCO}\equiv \frac{1}{\beta} \frac{m_1}{m_0}\,,
\end{equation}
which agrees with the results of Refs.~\cite{lrr-2006-3,Ohashi:1996uz}. If the small compact object is a black hole we get, by Eq.~\eqref{BHsensitivity}, $s_{\BH} = 1/2$. Finally, notice that Eq.~\eqref{gmunuJordan} reduces to Eq.~(2.11) in Ref.~\cite{Ohashi:1996uz}.

Brans-Dicke theory has already been strongly constrained by Solar System experiments. GR is recovered in the $\omega_{\BD} \to \infty$ limit and the tracking of the Cassini spacecraft has lead to the constraints $\omega_{\BD} > 4 \times 10^{4}$~\cite{Bertotti:2003rm}. 

\subsubsection*{Massive Brans-Dicke Limit}

Massive Brans-Dicke theory can be obtained with the same choices of coupling functions $F(\phi)$ and $Z(\phi)$ as in Eq.~\eqref{masslessBD}, which then leads to the same choice of $\alpha$ as in Eq.~\eqref{def_alpha}. Massive theories, however, require the addition of a non-vanishing potential
\be
U(\phi)=\frac{m_s^2}{2}(\phi-\phi_0)^2\,,
\label{massiveBD}
\ee
and hence, by Eq.~\eqref{mass_Jordan} $V_2=m_s^2\beta^2$ and
\begin{equation}
 \mu_s=\frac{m_s \beta}{2\sqrt{2\pi}}= \sqrt{\frac{2}{2 \omega_{\BD} + 3}} \; m_s\,.\label{def_mu}
\end{equation}
Therefore, massive Brans-Dicke theory is described by two-parameters only, $(\omega_{\BD},\mu_{s})$.


Before proceeding, notice that scalar-tensor theories comprise a class of models, of which Brans-Dicke theory is in principle a single member. However, to linear order in the mass ratio and assuming analyticity, we have shown that {\emph{all}} scalar tensor theories without (with) a potential reduce to massless (massive) Brans-Dicke. Therefore, any bound on the Brans-Dicke coupling parameter $\omega_{\BD}$ (or the Brans-Dicke mass $\mu_{s}$) is automatically a constraint on the coupling parameters of massless scalar-tensor theories (or the scalar mass) to linear order in the mass ratio. To next order in the mass ratio, this equivalence breaks down.

\section{EMRIs in the Teukolsky Framework}
\label{sec:GWST}

In this section, we describe the Teukolsky framework, which we use to study EMRIs in the test particle approximation. Hereafter, we restore the usual $G=1$ units. Consider then quasi-circular inspirals in the extreme mass-ratio limit: a small compact object with mass $m_{\SCO}$ and no spin orbiting around a supermassive black hole with mass $M_{\MBH}$, spin angular momentum $|S_{\MBH}^{i}| = a_{\MBH} M_{\MBH}$ and dimensionless Kerr spin parameter $q_{\MBH} = a_{\MBH}/M_{\MBH}$. 
We recall that $m_\SCO=A_0 m_0$, where $m_0$ is the bare mass.
The total mass of the system is then $M = m_{\SCO} + M_{\MBH}$, the reduced mass is $\mu_{\SCO} = m_{\SCO} M_{\MBH}/M \ll 1$ and the symmetric mass-ratio is $\nu = \mu_{\SCO}/M \ll 1/4$. We consider quasi-circular, equatorial EMRIs for simplicity. The spin of the small compact object can be neglected, as its effect on the evolution will be second-order in the mass ratio $\nu \ll 1$. Finally, in the extreme mass-ratio limit, possible tidal effects may also be neglected~\cite{Baumgarte:2004xq}.

With this at hand, let us consider the Teukolsky formalism~\cite{Teukolsky:1973ha}. In the test-particle limit, we can approximate the EMRI trajectories as geodesics of a test particle with mass $\nu$ is a background of mass $M$ to leading order in the mass-ratio. We can then study the gravitational and scalar waves emitted and the energy-momentum carried away by solving the first-order perturbation of the field equations as a function of the given geodesic. Notice that the scalar-tensor modified field equations~\eqref{EQBD1a}-\eqref{EQBD2a} are very similar to the Einstein equations with a propagating scalar field, a system originally studied by Detweiler~\cite{Detweiler:1978ge}. 

Any given geodesic is extremely sensitive to the background upon which it evolves, where we here choose the latter to be the Kerr metric. Sotiriou and Faraoni~\cite{Sotiriou:2011dz} recently showed that the most general, stationary, axisymmetric, vacuum spacetime that is also the endpoint of gravitational collapse in scalar-tensor theories is the Kerr metric. This does not apply to non-stationary backgrounds, with time-dependent hair of the form $\psi \sim \mu_{c} \; t$~\cite{2011arXiv1111.4009H}, where $\mu_{c}$ is a constant. In that case, $\dot{\psi} =\mu_{c}$ is assumed to be of cosmological origin, and thus much smaller than any frequency in the EMRI problem. 

Therefore, without loss of generality, we linearize Eq.~\eqref{EQBD1a}, the equations for the gravitational perturbations, about the Kerr background [in Boyer-Lindquist coordinates $(t,r,\theta,\phi)$] to first order in the mass ratio. One can decouple these equations in favor of a differential equation for the Newman-Penrose scalar $\psi_{4}$, which we can harmonically decompose via
\ba
\psi_{4} &=& \frac{1}{(r - i M_{\MBH} q_{\MBH} \cos{\theta})^{4}} 
\nonumber \\
&\times& \int_{-\infty}^{\infty} d \omega \sum_{l,m}R_{lm\omega}(r) \; S_{lm\omega}(\theta) e^{i(m \phi - \omega t)}\,,
\ea
where the radial function $R_{lm\omega}$ must satisfy the radial equation
\begin{equation}
 \Delta^2\frac{d}{dr}\Big(\frac{1}{\Delta}\frac{dR_{lm\omega}}{dr}\Big)-V_gR_{lm\omega}=-{\mathcal{T}}_{lm\omega}(r)\,,\label{radial}
\end{equation}
while the angular function $S_{lm\omega}$ must satisfy the angular equation
\begin{eqnarray}
 &&\frac{1}{\sin\theta}\frac{d}{d\theta}\Big(\sin\theta\frac{d}{d\theta}S_{lm\omega}\Big)-\Big(a_\MBH^2\omega^2\sin^2\theta+4a_\MBH\omega\cos\theta\nonumber\\
&&+\frac{m^2+4-4m\cos\theta}{\sin^2\theta}+2a_\MBH\omega m -2-\lambda\Big)S_{lm\omega}=0\,,\label{angular}
\end{eqnarray}
where $(l,m)$ are harmonic indices, $\lambda$ is the eigenvalue of Eq.~\eqref{angular} and
\begin{eqnarray}
\Delta&=&r^{2} - 2 M_{\MBH} r + M_{\MBH}^{2} q^{2}_\MBH,\nonumber\\
V_g&=&-\frac{1}{\Delta}\left[K^2-2i\frac{\Delta}{dr}K+\Delta\left(4i\frac{dK}{dr}-\lambda\right)\right],\nonumber\\
K&=&(r^2+a_\MBH^2)\omega+a_\MBH m\,,
\end{eqnarray}
The source term ${\mathcal{T}}_{lm\omega}$ can be derived from the stress-energy tensor of test-particles
\begin{equation}
 T^{\mu \nu}=\frac{m_\SCO u^{\mu} u^{\nu}}{\sqrt{-g} u^t}\delta(r-r_\SCO)\delta(\theta-\pi/2)\delta(\phi-\Omega_\orb t)\,,\label{stressparticle}
\end{equation}
where $r_\SCO$ is the location of the small compact object, $u^{\mu}$ is the small compact object's four-velocity and $\Omega_{\orb}$ is its orbital frequency. The source term and the details of how to solve Eqs.~\eqref{radial}-\eqref{angular} in terms of Green's functions are given in~\cite{Detweiler:1978ge}, so we omit them here.  

The equation for the evolution of the scalar perturbation in Eq.~\eqref{EQBD2a} can be solved by decomposing the latter via
\be
\varphi(t,r,\theta,\phi)=\sum_{l,m}\int d\omega e^{im\phi-i\omega t}\frac{X_{lm\omega}(r)}{\sqrt{r^2+ a_{\MBH}^2}} S_{lm}(\theta)\,,
\ee
where again $(l,m)$ are harmonic indices, $S_{lm}(\theta)$ are spheroidal harmonics and $X_{lm\omega}(r)$ is a radial function. With this parameterization, the scalar field evolution equation becomes
\be
\left[\frac{d^2}{dr_*^2}+V_s\right]X_{lm\omega}(r)=\frac{\Delta}{(r^2+a_{\MBH}^2)^{3/2}}T_{lm\omega}\,,\label{nonhom0}
\ee
where $dr/dr_*=\Delta/(r^2+a_{\MBH}^2)$ and the effective potential reads (see e.g. Ref.~\cite{Ohashi:1996uz})
\begin{eqnarray}
V_s&=&\left(\omega-\frac{a_{\MBH}m}{\hat\rho^2}\right)^2-\frac{\Delta}{\hat\rho^8}\left[\hat{\lambda}\hat\rho^4+2M_{\MBH}r^3+\right.\nonumber\\
&&\left.+a_{\MBH}^2(r^2-4M_{\MBH}r+a_{\MBH}^2)\right]-\frac{\Delta \mu_s^2}{\hat\rho^2}\,,
\end{eqnarray}
where $\hat\rho^2=r^2+a_{\MBH}^2$ and $\hat{\lambda}\equiv A_{lm}+a_{\MBH}^2\omega^2-a_{\MBH}^2\mu_s^2-2a_{\MBH}m\omega$ is found from the angular eigenfunction\footnote{Note that $\hat{\lambda}$ is different from the usual $\lambda$ defined, e.g. by Detweiler~\cite{Detweiler:1980uk} and Teukolsky~\cite{1973ApJ...185..635T} (in their notation,   
$\lambda=\hat{\lambda}+a_{\MBH}^2\mu_s^2$).}
\begin{eqnarray}
&&\frac{1}{\sin\theta}\frac{d}{d\theta}\left(\sin\theta\frac{dS}{d\theta}\right)+\nonumber\\
&&
\left[a_{\MBH}^2(\omega^2-\mu_s^2)\cos^2\theta-\frac{m^2}{\sin^2\theta}+A_{lm}\right]S=0\,.\nonumber
\end{eqnarray}
The scalar source function $T_{lm\omega}$ is different from that which sources gravitational perturbations, ${\cal{T}}_{lm\omega}$, and it is given by
\be
T_{lm\omega}=-\frac{\alpha}{u^t}S^*_{lm}\left(\frac{\pi}{2}\right)\delta(r-r_{\SCO})\,m_{\SCO}\delta(m\Omega_{\orb}-\omega)\label{tlmw}\,,
\ee
where the overhead star stands for complex conjugation. We recall that $\alpha$ is given by Eq.~\eqref{def_alpha} in the Brans-Dicke case.
 
Let us now consider two independent solutions $X_{lm\omega}^{r_+}$ and $X_{lm\omega}^{\infty}$ to the homogeneous version of Eq.~\eqref{nonhom0}, satisfying the following boundary conditions:
\be
X_{lm\omega}^{\infty,r_+}\sim e^{i r_* k_{{\infty},{+}} } \quad
{\rm as}
\quad
r\to \infty,\,r_+\,,\nonumber
\ee
where $k_+=\omega-m\Omega_+$, $\Omega_{+} = q_{\MBH}/(2 r_{+})$ and $k_{\infty}=\sqrt{\omega^2-\mu_s^2}$. The quantity $r_{+}$ is the largest root of $\Delta = 0$, namely $r_{\pm} = M_{\MBH} \pm M_{\MBH} \left(1 - q_{\MBH}^{2}\right)$, which is the location of the event horizon. Given this, the fluxes of scalar energy at the horizon and at infinity are~\cite{Teukolsky:1973ha}
\be
\dot E^{s}_{r_+,\infty}=\sum_{lm} m \; \Omega_{\orb} k_{+,\infty}|Z^{r_+,\infty}_{lm\omega}|^2\,,\label{scalar_fluxes}
\ee
where we have defined
\be
Z_{lm\omega}^{r_+,\infty}=-\alpha \, \nu \, \frac{X_{lm\omega}^{r_+,\infty}(r_{\SCO})}{W u^t}\frac{S_{lm}^{*}(\pi/2)}{\sqrt{r_{\SCO}^2+M_{\MBH}^{2}q_{\MBH}^2}}\,.
\label{Z}
\ee
and where $W$ is the Wronskian of the two linearly independent homogeneous solutions, $W=X_{lm\omega}^{r_+}{dX_{lm\omega}^{\infty}}/{dr_*}-X_{lm\omega}^{\infty}{dX_{lm\omega}^{r_+}}/{dr_*}$.

Through this formalism, we can then obtain the gravitational wave and scalar field energy flux to all orders in PN theory. This is because the Teukolsky approach employed here only linearizes the field equations in the mass-ratio, while retaining the full relativistic nature of the orbits. This approach, however, is only valid for closed geodesics, i.e.~provided the implicit averaging carried out through the harmonic decomposition is valid. This is sometimes referred to as the {\emph{adiabatic approximation}}, where one assumes the worldline of the small compact object can be approximated as a sequence of osculating geodesics. Such an approximation, of course, breaks down when the small compact object enters the ISCO, which is where we stop all our evolutions. Fortunately, the missing plunge and ringdown phases are completely negligible for data analysis purposes given EMRIs, since these two phases contribute negligible to the total signal-to-noise ratio.       

\section{Massless Scalar Energy Flux}
\label{Sec:EfluxMassless}

Let us first concentrate on massless scalar-tensor theories and compute the emitted scalar energy flux. Before solving the equations of Sec.~\ref{sec:GWST} numerically, let us describe certain analytical solution in the PN approximation, $v \ll1$ where $v$ is the binary's relative orbital velocity. In this limit, the scalar flux carried out to spatial infinity is~\cite{Ohashi:1996uz} 
\begin{eqnarray}
 {\cal{F}}^{s,\massless}_{\infty,\PN} \!\! &=&  \!\! \dot{{\cal{E}}}_{0}
 \left[1-2v^2+(2\pi-4q_\MBH)v^3
 \right. \nn\\
&+&  \!\! \!\!\left. (q_\MBH^2-10)v^4 +\left(\frac{12\pi}{5}+4q_\MBH\right)v^5 + {\cal{O}}(v^{6}) \right]\,.\nn\\
\label{dipole-massless}
\end{eqnarray}
with
\be
\dot{{\cal{E}}}_{0} \equiv \alpha^{2} \frac{M_{\MBH}^{2} m_{\SCO}^{2}}{12 \pi r^{4}}\,.
\ee
and $v=(M_\MBH\Omega_\orb)^{1/3}$. If one transforms to the Jordan frame via $m_{\SCO} = A_{0} m_{\SCO}^{(J)}$, $r = r^{(J)}/A_{0}$ with $A_{0}$ and $\alpha$ given by Eqs.~\eqref{def_A0} and~\eqref{def_alpha}, then the above agrees exactly with the results of Will and Zaglauer~\cite{1989ApJ...346..366W} and Ohashi, et al.~\cite{Ohashi:1996uz}.

The fully-relativistic, scalar energy flux computed in the Teukolsky approach from the numerical solution to the equations presented in Sec.~\ref{sec:GWST} reads
\begin{equation}
 {\cal{F}}_{\Tot}^{s,\massless}=2\sum_{l=1}^\infty\sum_{m=0}^l\left[m\Omega_{\orb}\left( k_{+}|Z^{r_+}_{lm\omega}|^2+k_{\infty}|Z^{\infty}_{lm\omega}|^2\right)\right]\,,\label{series}
\end{equation}
Notice that the $l = 1$ contribution to the sum does not here vanish, due to the presence of dipole radiation. Notice also that there are two contributions: one that escapes to spatial infinity and one that is absorbed by the event horizon of the supermassive black hole. Although the latter is usually small relative to the former, and thus it is mostly neglected in the literature, we here include it. When numerically solving Eq.~\eqref{scalar_fluxes} for a given $q_\MBH$, we truncate the sum in $l$ when the series evaluated at the ISCO, $r_\SCO=r_\ISCO$, converges to one part in $10^5$ or better. This requires summing up to $l=17$ for $q_\MBH=0.99$, but only up to $l=6$ for $q_\MBH=0$. Such a scheme then implies that our numerical data is accurate to one part in $10^{5}$, which is sufficient for this study. 

\begin{figure}[htb]
\epsfig{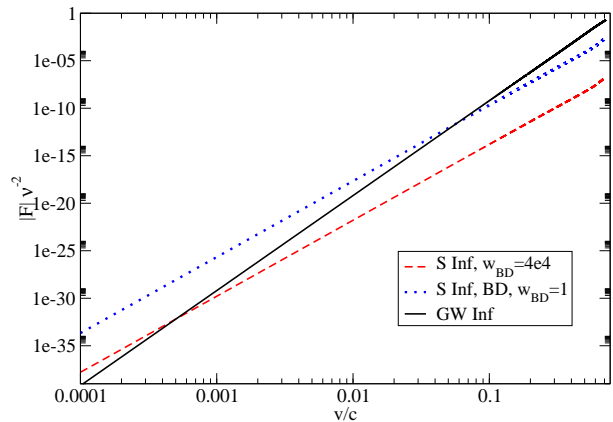}
\caption{\label{fig:EdotTot} PN expansion of the ($\nu^{2}$-normalized) gravitational wave energy flux (black solid curve) and the scalar energy flux carried out to infinity for Brans-Dicke theory with $\omega_{\BD} = 4 \times 10^{4}$ (dashed red curve) and $\omega_{\BD} = 1$ (dotted blue curve) as a function of velocity for a supermassive black hole with spin $q = 0.99$ and a small compact object with $s_{\SCO} = 0.188$. Observe that the scalar flux dominates over the GR one at sufficiently small velocities.}
\end{figure}
With this at hand, one might wonder how the gravitational wave flux compares to this scalar flux. Figure~\ref{fig:EdotTot} plots the PN expansion of the ($\nu^{2}$-normalized) gravitational wave energy flux (black solid curve) and the scalar energy flux carried out to infinity with $\omega_{\BD} = 4 \times 10^{4}$ (dashed red curve) and $\omega_{\BD} = 1$ (dotted blue curve) as a function of velocity for a supermassive black hole with spin $q_\MBH = 0.99$. Observe that the scalar flux becomes larger than the GR gravitational wave flux at sufficiently small velocities. Saturating the Solar system bound ($\omega_{\BD} > 4 \times 10^{4}$), the gravitational wave energy flux dominates for velocities $v > v_{\rm th}$, where $v_{\rm th} \sim 10^{-3}$. Notice, however, that the {\emph{difference}} between the gravitational wave and scalar fluxes is smallest close to $v_{\rm th}$. That is, gravitational waves in massless scalar-tensor theories are the most different from GR waves for $v \lesssim v_{\rm th}$, far away from the ISCO. 

\begin{figure*}[htb]
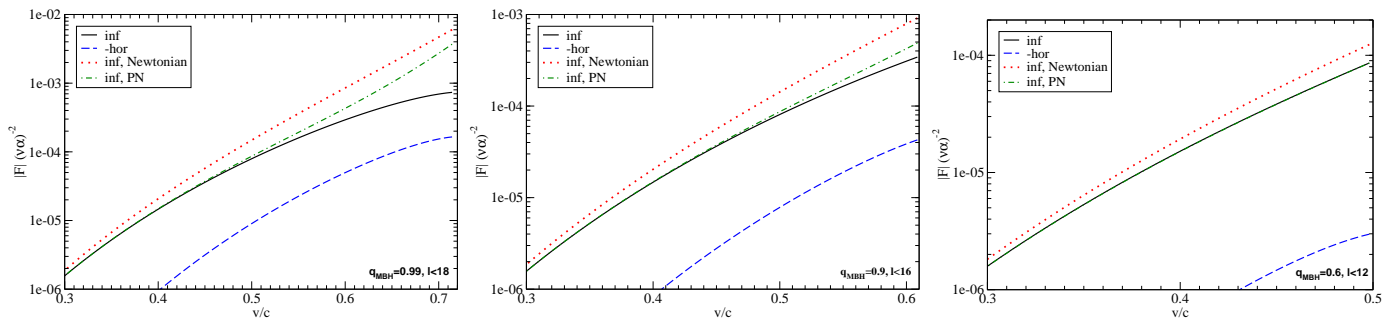

\begin{center}
\begin{tabular}{ccc}
\epsfig{file=scalar_fluxes_a099.eps,width=6cm,angle=0,clip=true}&
\epsfig{file=scalar_fluxes_a09.eps,width=6cm,angle=0,clip=true}&
\epsfig{file=scalar_fluxes_a06.eps,width=6cm,angle=0,clip=true}
\end{tabular}
\caption{\label{fig:scalar_fluxes} Scalar flux at infinity (black solid curve) and at the horizon (blue dashed curve) for a generic massless scalar-tensor theory, normalized by $(\alpha \nu)^2$ as functions of the velocity $v$ for $\mu_s=0$, $q_{\MBH}=0.99$ (left panel), $q_{\MBH}=0.9$ (middle panel) and $q_{\MBH}=0.6$ (right panel). For comparison, we show the PN formula~\eqref{dipole-massless} (green dot-dashed line) and its truncation at $-1$PN order (red dotted line).}
\end{center}
\end{figure*}
Let us now compare the analytic expression for the scalar flux to the numerical one. Figure~\ref{fig:scalar_fluxes} shows the numerical energy flux -- normalized by $(\alpha \nu)^2$ -- carried to infinity ${\cal{F}}^{s,\massless}_{\infty}$ (black solid) and absorbed by the supermassive black hole ${\cal{F}}^{s,\massless}_{+}$ (dashed blue), as well as its leading-order (Newtonian) approximation ${\cal{F}}_{\infty,\Newt}^{s,\massless}$ (dotted red)~\cite{1989ApJ...346..366W} and its higher-order PN expansion ${\cal{F}}^{s,\massless}_{\infty,\PN}$ (dot-dashed green)~\cite{Ohashi:1996uz} as a function of velocity. First, observe that the flux absorbed by the horizon is always at least an order of magnitude smaller than that carried out to infinity. Second, observe that as the small compact object approaches the ISCO, the full numerical results deviate from the analytic approximations. The disagreement is larger in the near-extremal case, although even then the $2.5$PN approximation of Eq.~\eqref{dipole-massless} does remarkably well. For non-spinning or slowly-rotating supermassive black holes, even the leading order approximation reproduces the exact results to high accuracy.

For corotating orbits, the effect of the high-order PN terms absent from Eq.~\eqref{dipole-massless} is to {\emph{reduce}} the overall magnitude of the flux, so that the PN formula \emph{overestimates} the exact flux. In fact, the leading-order, Newtonian flux formula can overestimate the flux by more than an order of magnitude for orbits close to the ISCO of a maximally spinning black hole. Moreover, for highly spinning black holes, the horizon flux also contributes to reduce the total flux (for $q_\MBH \gtrsim 0.36$, any orbit outside the ISCO satisfies the superradiant condition, so that the scalar flux at the horizon is \emph{negative}, cf. Fig.~\ref{fig:isco} later presented). On the other hand, for counter-rotating orbits, the numerical flux tends to be larger than the corresponding PN formula. Since in this case the ISCO is located far away from the horizon ($6 M_\MBH\leq r_\ISCO<9M_\MBH$ for $0\leq q_\MBH<1$), the deviations from the PN formula are smaller.

\begin{table}
 \begin{tabular}{c|cc}
    &	Fit I	&	Fit II	\\
\hline
$a_0$		&12.613	  &0.15240	\\
$a_1$		&-30.520  &-19.619	\\
$a_2$		&13.742	 &7.8369	\\
$a_3$		&-1.2497  &-6.5969	\\
$a_4$		&0		 &16.569	\\
$a_5$		&0	  	 &-21.100	\\
$a_6$		&0		 &13.959	\\
$a_7$		&0		 &-1.4791	\\
$a_8$		&0		 &0	\\
$a_9$		&0		 &0	\\
\hline
$1-R^2$ & $6 \times 10^{-6}$	&  $3 \times10^{-8}$	
 \end{tabular}
\caption{\label{tab:fit} Coefficients of the extra term~\eqref{eq:fit} in the dipole formula~\eqref{dipole-massless} obtained by fitting our numerical flux. As a measure of the fit error, we present the deviation of the coefficient of determination, $R^2$, from unity.}
\end{table}
Our numerical results can be also used to estimate some higher-order PN terms missing from Eq.~\eqref{dipole-massless}. We have considered the following corrections to the expression in square brackets in Eq.~\eqref{dipole-massless}:
\ba
{\cal{F}}^{s,\massless}_{\infty} \!\! &=& \!\! {\cal{F}}^{s,\massless}_{\infty,{\rm{PN}}} +  \dot{{\cal{E}}}_{0} \left[a_0+a_1 q_{\MBH}+a_2 q_{\MBH}^2
\right. 
\nonumber \\
&+& \! \! \! \! \left.
a_3\log\left(v\right)\right]v^6 + \left[a_4+a_5 q_{\MBH}+a_6 q_{\MBH}^2
\right. 
\nonumber \\
&+& \! \! \! \! \left.
a_7 q_{\MBH}^3+ a_8 q_{\MBH}^4+a_9\log\left(v\right)\right]v^7\,.
\label{eq:fit}
\ea
The structure of this fitting function is based on the $3$PN and $3.5$PN structure of the energy flux expansion in GR. The coefficients $a_i$ can be computed by fitting our numerical data, obtained spanning the region $q_\MBH\in\left[-0.99,0.99\right]$ and $v\in\left[0.01,v_\ISCO\right]$. We attempted approximately a hundred different fits, varying the number of fitting parameters, the number of PN order corrections, and the accuracy of the data to be fitted. The best fits we found, ie.~those that provide the smallest errors with the least number of additional parameters, are shown in Table~\ref{tab:fit} for two different fits. Notice that these parameters are not meant to represent the next order terms in the PN series, but they are just a fit to our numerical data to improve the analytical representation of the function.

\begin{figure}[htb]
\epsfig{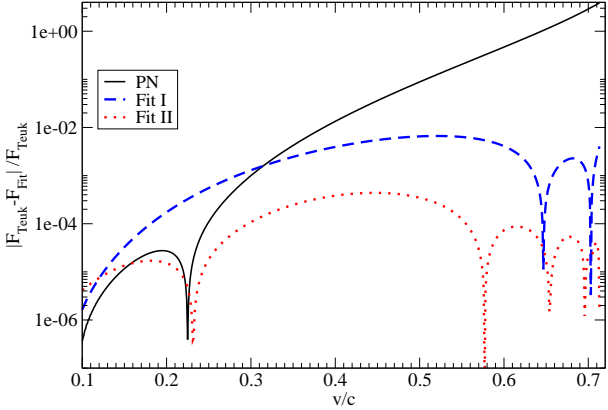}
\caption{\label{fig:diff_fit} Fractional difference between the fits, ${\cal{F}}^{s,\massless}_{\infty}$, with parameters from Table~\ref{tab:fit} and Teukolsky-based fluxes , ${\cal{F}}_{\Teuk}$, as a function of velocity for spin $q_\MBH=0.99$ and compared with the fractional difference with the PN flux~\eqref{dipole-massless}. These curves are an upper bound of the fit error: the fractional difference for Fit I, Fit II and for the PN formula decreases for smaller values of $q_\MBH$.}
\end{figure}

Figure~\ref{fig:diff_fit} shows the fractional difference between the numerical flux and Fit I (blue dashed curve) or Fit II (dotted red curve), which should be compared with the difference between the numerical flux and the bare PN result of Eq.~\eqref{dipole-massless} (black solid curve). This figure presents results for a supermassive black hole with spin $q_\MBH=0.99$ as a function of velocity. While the PN formula can deviate by a factor $300\%$ close to the ISCO, Fits I and II introduces errors smaller than $1\%$ and $0.02\%$, respectively. Numerical inaccuracy of our Teukolsky-based flux at small velocity ($v\lesssim0.1$) prevents us from obtaining reliable fits that deviate less than 1 part in $10^4$. Note that the fractional difference shown in Fig.~\ref{fig:diff_fit} represents an upper bound; the errors of our fits decrease for less extreme values of $q_\MBH$. This is because the PN flux of Eq.~\eqref{dipole-massless} agrees better with the numerical flux for lower values of $q_{\MBH}$ and because the ISCO velocity decreases with decreasing $q_{\MBH}$.  

Finally, let us consider the effect of this higher-order PN terms in the flux on the Fourier transform of the response function in the stationary-phase approximation. That is, we solve a generalized Fourier integral assuming that the gravitational phase oscillates much more rapidly than its amplitude (see eg.~\cite{Droz:1999qx,Yunes:2009yz}). After averaging over beam-pattern functions, the Fourier transform of the response can be written as
\be
\tilde{h}(f) = {\cal{A}} \frac{{\cal{M}}^{5/6}}{D_{L}} f^{-7/6} e^{i \Psi(f)}
\ee 
where ${\cal{A}}$ is a constant amplitude, ${\cal{M}} = \nu^{3/5} M$ is the chirp mass, $D_{L}$ is the luminosity distance, and $f$ is the gravitational wave frequency. The Fourier phase can be computed from
\be
\Psi(f) = 2 \pi \int^{f/2} \frac{F'}{\dot{F}'_{\orb}} \left(2 - \frac{f}{F'}\right) dF'\,,
\ee
where $\dot{F}_{\orb}$ is the rate of change of the orbital frequency, which can be computed by the chain rule: $\dot{F}_{\orb} = \dot{E}_{b} (dE_{b}/dF_{\orb})^{-1}$. The quantity $\dot{E}_{b}$ is the rate of change of the binding energy, which from the balance law satisfies $\dot{E}_{b} = -{\cal{F}}_{\Tot}$, where ${\cal{F}}_{\Tot}$ is the total energy flux, while $dE_{b}/dF$ can be obtained by differentiating the binding energy $E_{b}$, given later in Eq.~\eqref{E-binding}. The total energy flux ${\cal{F}}_{\Tot} = {\cal{F}}_{\GR}^{g} + {\cal{F}}^{s,\massless}_{\Tot}$, where ${\cal{F}}_{\GR}^{g} = (32/5) \nu^{2} v^{10} [1 + {\cal{O}}(v^{2})]$ is the gravitational energy flux in GR, while we will model ${\cal{F}}^{s,\massless}_{\Tot} =  {\cal{F}}^{s,\massless}_{\infty,\PN}$.  We should note here that we have neglected the contribution to the rate of change of the binding energy due to horizon absorption effects. These will enter at ${\cal{O}}(v^{5})$ relative to the leading-order, Newtonian effect (see e.g.~\cite{Mino:1997bx}). We here only concentrate on high PN order corrections induced by the energy flux carried out to spatial infinity, such that horizon absorption effects can be systematically included in the future if needed.

Using this and performing the integration, the Fourier phase becomes $\Psi(f) = \Psi_{\GR} (f) + \delta \Psi$, where the GR Fourier phase is
\ba
\Psi_{\GR}(f) &=& 2 \pi f t_{c} - \phi_{c}  + \frac{3}{128} u^{-5/3} \left[ 1 
\right. 
\nonumber \\
&+& \left. \left(\frac{3715}{756} + \frac{55}{9} \nu \right) \nu^{-2/5} u^{2/3} + {\cal{O}}(u) \right],
\ea
with $(t_{c},\phi_{c})$ the time and phase of coalescence and $u \equiv \pi {\cal{M}} f$ the reduced gravitational wave frequency. Higher-order terms in the GR Fourier phase in the PN approximation can be found eg.~in~\cite{Berti:2005qd}. The Brans-Dicke correction is
\ba
\delta \Psi(f) &=& -\frac{5 \left(s_{\SCO} - \frac{1}{2} \right)^{2}}{3584  \left(\omega_{\BD} + 2 \right)}  
u^{-7/3} \nu^{2/5} \left[ 1 - \frac{7}{2} \nu^{-2/5} u^{2/3} 
\right. 
\nonumber \\
&+& \left. 
\left(5 \pi - 10 q_{\MBH}\right) \nu^{-3/5} u
+ \left(\frac{35}{9} q_{\MBH}^{2} - \frac{350}{9} \right) 
\right. 
\nonumber \\
&\times& \left. \nu^{-4/5} u^{4/3} 
 +\left(\frac{84}{5} \pi + 28 q_{\MBH} \right) \nu^{-1} u^{5/3} 
 \right. 
 \nonumber \\
 &+& \left. {\cal{O}}(u^{2}) \right]\,.
 \label{SPA-phase}
\ea
Notice that the leading-order correction agrees identically with the results of~\cite{Will:1994fb} in the large $\omega_{\BD}$ limit, while the higher PN terms are new. Notice also that the Fourier phase presents similar features to the energy flux. That is, the higher PN order terms seem to counteract the leading-order, Newtonian term, leading to a smaller number of additional gravitational wave cycles induced by the Brans-Dicke correction. This then implies that calculations obtained by using the leading-order, Newtonian term only will tend to {\emph{overestimate}} the effect of massless Brans-Dicke theory. 

One can now map the deformed Fourier phase to the ppE framework~\cite{Yunes:2009ke}. In the latter, one postulates a waveform family of the form $\tilde{h} = \tilde{h}_{\GR} \delta \tilde{h}$, where $\tilde{h}_{\GR}$ is the GR Fourier transform of the response function, while $\delta \tilde{h}_{\ppE}$ is a parametric deformation. The latter can be written as  $\delta \tilde{h}_{\ppE} = (1 +  \delta A_{\ppE}) \; \exp(i \delta \Psi_{\ppE})$, where $\delta A_{\ppE}$ and $\delta \Psi_{\ppE}$ are amplitude and phase corrections that depend on frequency. In its simplest incarnation, one can expand $\delta A_{\ppE} = \alpha u^{a}$ and $\delta \Psi_{\ppE} = \beta u^{b}$, i.e.~as a parametric power series in $u$, where $(\alpha,\beta,a,b)$ are ppE parameters. Clearly, in our case the simplest realization will not do; one must use a more general parameterization, such as Eq.~$(45)$ in~\cite{Yunes:2009ke}, namely $\delta A_{\ppE} = 0$ and  
\be
\delta \Psi_{\ppE} = \sum_{k=0}^{K} \phi_{k} u^{(k-7)/3}\,.
\ee
To match the high-PN order expansion of the phase in Eq.~\eqref{SPA-phase}, we need five terms with the mapping 
\ba
\phi_{0} &=&  -\frac{5 \left(s_{\SCO} - \frac{1}{2} \right)^{2}}{3584  \left(\omega_{\BD} + 2 \right)}  
 \nu^{2/5}\,,
\qquad
\phi_{1} = 0\,,
\nonumber \\
\phi_{2} &=& - \frac{7}{2} \nu^{-2/5} \phi_{0}\,,
\nonumber \\
\phi_{3} &=&\left(5 \pi - 10 q_{\MBH}\right) \nu^{-3/5} \phi_{0}\,,
\nonumber \\
\phi_{4} &=& \left(\frac{35}{9} q_{\MBH}^{2} - \frac{350}{9} \right)  \nu^{-4/5} \phi_{0}\,,
\nonumber \\
\phi_{5} &=& \left(\frac{84}{5} \pi + 28 q_{\MBH} \right) \nu^{-1} \phi_{0}\,.
\ea
We see then that the next-to-simplest incarnation of the ppE framework easily manages to map to the high-PN order Brans-Dicke prediction for the Fourier phase. 

\section{Massive Scalar Energy Flux}
\label{Sec:EfluxMassive}

Let us now concentrate on massive scalar-tensor theories. First, let us consider the scalar field energy flux carried out to infinity by a massive scalar field. Solving Eq.~\eqref{EQBD2a} in the large distance limit, one obtains
\be
{\cal{F}}^{s,\massive}_{\infty}=\frac{\alpha^2 M_{\MBH}^4}{12\pi r_{\SCO}^4}
\left(1-\mu_s^2 \frac{r_{\SCO}^3}{M_{\MBH}}\right)^{3/2}
\nu^2 {\cal H}(\Omega_{\orb}-\mu_s)\,,
\label{dipole}
\ee
where ${\cal H}(x)$ is the Heaviside function and we have only accounted for the dominant mode. Notice that this flux is non-vanishing only for masses that satisfy $\Omega_{\orb} - \mu_{s} > 0$, which implies $M_{\MBH} \mu_{s} < v^{3}$. For typical EMRI velocities, $v \sim 0.2$, the scalar field energy flux carried out to infinity is then non-vanishing only if $\mu_{s} <  10^{-2}/M_{\MBH}$. For larger masses, the scalar mode cannot overcome the massive potential barrier and it becomes trapped. These results can be generalized to an arbitrary harmonic mode $m$ via $\Omega_\orb<\mu_s/m$, or equivalently $M_{\MBH} \mu_{s} < m v^{3}$.

\begin{figure}[htb]
\begin{center}
\begin{tabular}{cc}
\epsfig{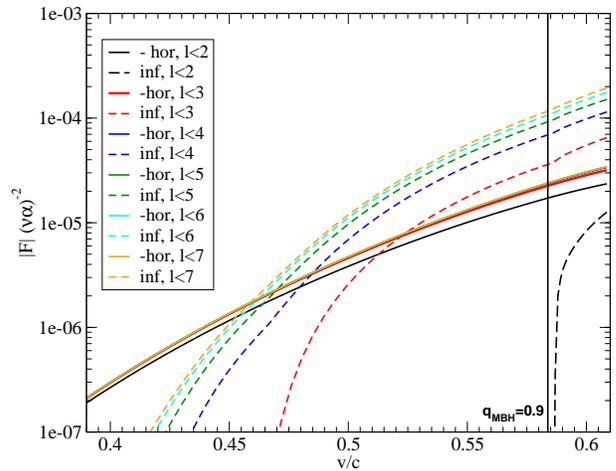}
\end{tabular}
\caption{\label{fig:scalar_fluxes_massive} Scalar energy flux, normalized by $(\alpha \nu)^2$, carried to infinity and into the supermassive black hole horizon as a function of the velocity for $\mu_s M_\MBH=0.2$ and $q_\MBH=0.9$. Different curves are obtained by truncating the series in Eq.~\eqref{series} at different values of $l$. Observe that below a certain velocity, $v\lesssim0.4$, the contribution to infinity vanishes in accordance with Eq.~\eqref{dipole}. The resonance corresponding to $l=m=1$ is also shown.}
\end{center}
\end{figure}

Figure~\ref{fig:scalar_fluxes_massive} shows the scalar flux carried to infinity and that absorbed by the supermassive black hole with $q_\MBH=0.9$, $\mu_s M_\MBH=0.2$ and for different values of $l$ at which the series of Eq.~\eqref{series} is truncated. Observe that the contribution to the scalar flux at infinity vanishes at the orbital distance that corresponds to $\Omega_\orb=\mu_s/m$, in agreement with Eq.~\eqref{dipole}. Hence, at sufficiently large orbital separation, the contribution from the flux at the horizon will eventually become dominant, as it occurs at $v \lesssim 0.45$ in Fig.~\ref{fig:scalar_fluxes_massive}. This figure also shows the rate of convergence of the scalar energy flux as a function of number of $l$ modes summed over. Observe that although the flux absorbed by the horizon converges rather fast, the one carried out to infinity requires a large number of $l$ modes. Overall, however, the total massive scalra flux is generically smaller than the massless scalar flux (compare Fig.~\ref{fig:scalar_fluxes_massive} with the middle panel of Fig.~\ref{fig:scalar_fluxes}). If this were the whole story, we would therefore generically expect smaller scalar-tensor corrections in the massive case. 

Massive scalar-tensor theories, however, have the most interesting feature of generically leading to certain resonances in the part of the scalar flux that impinges on the supermassive black hole horizon, as recently studied in~\cite{Cardoso:2011xi}. Figure~\ref{fig:scalar_fluxes_massive} shows such a resonance at $v\sim0.5838$. These resonances occur when the orbital frequency excites the quasinormal frequencies of oscillation of the scalar field, which are approximately proportional to the scalar field mass~\cite{Cardoso:2005vk,Dolan:2007mj,Berti:2009kk}. In the limit $\mu_{s} \ll 1/M_{\MBH}$, the resonant frequency is~\cite{Detweiler:1980uk}:
\be
\Omega_{\res}=\mu_s \left[1- \left(\frac{\mu_s M_{\MBH}}{l+1+n}\right)^2\right]^{1/2}\,,
\quad n=0,1,...
\label{om_resonance}
\ee
Clearly then, $\Omega_{\res} \sim \mu_{s}$ in the $\mu_{s} M_{\MBH} \ll 1$ limit. We can relate this to the orbital radius via
\begin{equation}
r=r(\Omega_\orb)= {M_{\MBH}} \frac{\left[1 - q_\MBH\, (M_{\MBH} \Omega_{\orb}) \right]^{2/3}}{(M_{\MBH} \Omega_{\orb})^{2/3}}\,,
\label{r-of-w}
\end{equation}
where we define the resonance location as $r_\res=r(\Omega_\res)$. Obviously, these resonances are not present in the massless case as then $\Omega_\res \to 0$.

Such resonances greatly enhance the magnitude of $Z_{lm\omega}^{r_{+},\infty}$ and, in turn, the amount of energy flux radiated~\cite{Cardoso:2011xi}. In Appendix~\ref{app:resonances}, we compute the resonant flux analytically in the $\mu_s M_\MBH\ll1$ limit, for any $(l,m)$ and for the fundamental mode (cf. Eq.~\eqref{res-height_gen}). When $l=m=1$, Eq.~\eqref{res-height_gen} reduces to 
\ba
{\cal{F}}^{s,\massive}_{+} &\sim& -\frac{3 \alpha ^2}{16 \pi} \nu^2 \frac{M_{\MBH}}{r_{+}} \sqrt{\f{r_\res}{M_\MBH}} \left(1-q_\MBH^2\right)^{-1}  {\cal B}^{-1}
\nonumber \\
&\times&
\left[\frac{q_\MBH}{2} \frac{M_\MBH}{r_+}-\left(\frac{M_\MBH}{r_\res}\right)^{3/2} \right]^{-1}\,,
\label{res-height}
\ea
where ${\cal B}=1+4 {\cal K}^2$ and ${\cal K}=-2M_{\MBH} r_+ k_+/(r_+-r_-)$. We have found very good agreement between this expression and our numerical solutions for $\mu_s M_\MBH\lesssim0.1$. 

In the small $\mu_s$ limit, the resonance location scales as $r_\res\sim\mu_s^{-2/3}$ and the scalar flux at resonance \emph{grows} in magnitude when $\mu_s\to0$. For generic $l$ and $q_{\MBH}\neq0$, the peak flux scales as ${\cal{F}}^{s,\massive}_{+,{\rm peak}}\sim \mu_s^{1-4l/3}$ [cf.~Eq.~\eqref{res-height_gen} in Appendix~\eqref{app:resonances}]. For very small $q_{\MBH}$, the peak flux at resonance is instead positive, and it can also be very large. For example for a Schwarzschild black hole, 
\be
{\cal{F}}^{s, \massive}_{+,{\rm peak}}\sim  \frac{3\alpha ^2\nu^2}{32 \pi }(\mu_s M_\MBH)^{-4/3} \,.                                                                                                                                                                                                                                                                                                                                                                                                                                                                                                                                                                                                                                                                                                                                                                                                                                                                                                                                                                                                                                                                     
\ee

In the small $\mu_s$ limit, the resonance width is twice the imaginary part of the quasinormal mode frequency~\cite{Detweiler:1980uk}
\ba
\Delta \Omega^{lmn} &=&\mu_s (\mu_s M_{\MBH})^{4l+4}\left(q_{\MBH} m-2 \mu_s r_+\right) 
\nonumber \\
&\times&
\frac{2^{4l+2}(2l+1+n)!}{n!(l+1+n)^{2l+4}} \left[\frac{l!}{(2l)!(2l+1)!}\right]^2
\nn\\
&\times&
\prod_{j=1}^l\left[j^2\left(1-q^2_\MBH\right)+\left(mq_\MBH-2 \mu_s r_+\right)^2\right]\,.\nn\\\label{width}
\ea
For the fundamental mode, $n=0$ with $l=m=1$, and in the small $\mu_s$ limit, the equation above reduces to
\begin{equation}
 \Delta\Omega \sim \frac{1}{12 M_\MBH} \left(\mu_s M_\MBH\right)^9 \left(q_\MBH-2r_+\mu_s \right) \,.
 \label{width-simplified}
\end{equation}
Notice that $M_{\MBH}\Delta \Omega \propto (\mu_{s} M_{\MBH})^{9}$ and it is thus incredibly small for small $\mu_{s} M_{\MBH}$. For larger values of $\mu_s M_\MBH$, the width can be computed numerically, using  a continued fraction method~\cite{Cardoso:2005vk,Dolan:2007mj,Berti:2009kk}. Some numerical results are shown in Table~\ref{tab:resonance99}, where $\Delta R=\Delta\Omega\partial r /\partial \Omega$ is the resonance width evaluated at $r=r_\res$ using Eq.~\eqref{r-of-w}. 

\begin{center}
\begin{table}
\begin{tabular}{ccccc}
\hline
\hline
$\mu_s M_\MBH$ & $\,\,\,r_\res/M_\MBH$ & $\,\,\,(\alpha\nu)^{-2}\dot E^{s,\,{\rm peak}}_{r_+}$ &	 $\Delta R/(2M_\MBH)$\\
\hline 
$0.35$	&	$1.52818949793075$	  &	$-0.0522$	                      &	$1.15\times10^{-7}$\\
$0.3$	&	$1.78503938021340$	  &	$-0.0752$	                      &	$1.52\times10^{-7}$\\
$0.25$	&	$2.10016240393323$	  &	$-0.0969$	                      &	$5.93\times10^{-8}$\\
$0.2$	&	$2.53500275855866$	  &	$-0.1200$	                      &	$1.82\times10^{-8}$\\
$0.15$	&	$3.18939434130550$	  &	$-0.1467$	                      &	$3.30\times10^{-9}$\\
$0.1$   &	$4.33400288873563$	  &	$-0.1828$	              	      &	$4.38\times10^{-10}$\\
\hline
\hline
\end{tabular}
\caption{\label{tab:resonance99} 
Orbital radius at resonance and peak scalar flux for $n=0$, $l=m=1$, $q_\MBH=0.99$ and several values of $\mu_s M_\MBH$. $\Delta R$ is the resonance width in the radial dimension.}
\end{table}
\end{center}

Scalar flux resonances can lead to certain {\emph{floating orbits}}, as recently studied in~\cite{Cardoso:2011xi}. Such floating orbits are defined as those where the scalar flux identically cancels the ${\cal{O}}(\nu^{2})$ gravitational wave flux, thus greatly slowing down the inspiral. The inspiral evolution is then sourced by the shedding of mass and angular momentum of the supermassive black hole. Cardoso et al.~\cite{Cardoso:2011xi} found that such orbits are possible outside the ISCO, provided $q\MBH\gtrsim0.36$ and the scalar mass $\mu_s$ is small enough. Section~\ref{sec:Constraints-Massive} will consider how such resonances can be used to constrain massive scalar-tensor theories. 

\section{Projected Constraints on Massless Scalar-Tensor Theories}
\label{sec:Constraints}

In this section we will study the constraints one could place on massless scalar-tensor theories given a gravitational wave observation of an EMRI. We begin with a review of the EOB approach~\cite{Damour:1997ub,Buonanno99,Buonanno00,Damour00,Damour01,Buonanno06,Nagar:2006xv,Damour2007,Damour:2008qf,Damour:2008gu,Damour:2009kr,Buonanno:2009qa,Barausse:2009xi,Pan:2009wj,Barack:2009ey,Damour:2009sm,Bernuzzi:2010ty,Pan:2010hz,Fujita:2010xj,Fujita:2011zk,Nagar:2011fx,Barausse:2011ys}, following in particular the model developed by Yunes, et al.~\cite{Yunes:2009ef,2009GWN.....2....3Y,Yunes:2010zj}. We then continue with a discussion of how this model can be modified to include scalar-tensor modifications. We conclude with a numerical evolution of orbits in the EOB-EMRI scheme to determine plausible projected constraints. Note that, when needed, we shall map the metric perturbations from the Einstein frame to the Jordan frame, by using Eq.~\eqref{gmunuJordan}. 

\subsection{EOB Modeling of EMRIs in General Relativity}
\label{sec:EMRIMod}

The use of the EOB-EMRI approximation is not fundamental to this paper. First, the only EMRI orbital regime of interest for data analysis is the inspiral, as almost all of the signal-to-noise ratio builds during this regime. Therefore, the EOB-EMRI approach does not account for the plunge and merger, as these phases are irrelevant for EMRIs. Second, we choose an EOB-EMRI model because modifying it to include scalar-tensor theory corrections is fast and straightforward. We could have instead used a time-domain Teukolsky evolution~\cite{2007PhRvD..76j4005S,2008PhRvD..78b4022S}, or an interpolated frequency-domain Teukolsky evolution~\cite{Hughes:1999bq,Hughes:2001jr}. However, in Refs.~\cite{Yunes:2009ef,2009GWN.....2....3Y,Yunes:2010zj}, it has been shown that an EOB-EMRI approach is equally valid during the inspiral phase up to the ISCO for quasi-circular orbits in the equatorial plane.

The EOB-EMRI framework makes use of the so-called adiabatic approximation, i.e.~the radiation-reaction time-scale is much longer than the orbital one. In this approximation, the gravitational wave phase can be obtained by solving
\begin{eqnarray}
\label{omegaadiab}
\dot{\Omega}_{\orb} &=& \left(\frac{d E_{b}}{d \Omega_{\orb}}\right)^{-1} \, \dot{E}_{b}(\Omega_{\orb})\,,
\quad
\dot{\phi} = \Omega_{\orb}\,, 
\end{eqnarray}
where $\Omega_{\orb}$ is obviously the small compact object's orbital angular frequency and $\phi$ is the orbital phase, $E_{b}$ is the system's binding energy and $\dot{E}_{b} = -{\cal{F}}_{\Tot}$ is the total rate of change of this energy, where ${\cal{F}}_{\Tot}$ is the energy flux. 

We model the binding energy of the system as that of a test-particle in orbit at radius $r$ around a Kerr black hole~\cite{Bardeen:1972fi}:
\begin{equation}
E_{b}=m_{\SCO}\,\frac{1-2 M_{\MBH}/r_\SCO \pm q_{\MBH}\,M_{\MBH}^{3/2}/r_\SCO^{3/2}}{\sqrt{1-3 M_{\MBH}/r_\SCO\pm 2q_{\MBH}\,M_{\MBH}^{3/2}/r_\SCO^{3/2}}}\,.
\label{E-binding}
\end{equation}
where the $\pm$ stands for prograde or retrograde orbits. This is the exact Hamiltonian for a test-particle in orbit around a Kerr black hole (without the constant rest mass piece). Therefore, Hamilton's equations are equivalent to the evolution of geodesics in the Kerr spacetime.  

In GR, the only energy sink is that induced by the emission of gravitational waves. Because of this, one assumes the balance law $\dot{E}_{b} = - {\cal F}^{g}_{\GR,\Tot}$, where the latter is the total energy flux ${\cal{F}}^{g}_{\GR,\Tot} = {\cal{F}}^{g}_{\GR,\infty} + {\cal{F}}^{g}_{\GR,+}$. The component that describes radiation that escapes to spatial infinity is labeled ${\cal{F}}^{g}_{\GR,\infty}$, while the one that describes radiation absorbed by the supermassive black hole is labeled ${\cal{F}}^{g}_{\GR,+}$. The former can be approximated as~\cite{Damour2007,Damour:2008gu,Pan:2010hz}
\begin{eqnarray}
{\cal{F}}^{g}_{\GR,\infty}(\Omega_{\orb})= \frac{1}{8 \pi}\,\sum_{\ell=2}^{8}\sum_{m=0}^{\ell} (m\,\Omega_{\orb})^2\,\left|h_{\ell m} \right|^{2}\,,
\label{pd-flux-adiab}
\end{eqnarray}
where the multipole-decomposed gravitational wave can be re-expressed in the following factorized form 
\begin{equation}
h_{\ell m}(v) = h_{\ell m}^{\Newt,\epsilon_p}\;S^{\epsilon_p}_{\ell m} \; T_{\ell m}\; e^{i \delta_{\ell m}}\; (\rho_{\ell m})^\ell\,.
\label{full-h}
\end{equation}
Here, $\epsilon_p$ is the parity of the waveform (\textit{i.e.,} $\epsilon_p=0$ if $\ell+ m$ is even, $\epsilon_p=1$ if $\ell+ m$ is odd), while $S^{\epsilon_p}_{\ell m}(v)$, $T_{\ell m}(v)$, $\delta_{\ell m}(v)$ and $\rho_{\ell m}(v)$ can be found in~\cite{Damour2007,Damour:2008gu,Pan:2010hz}.  The Newtonian waveform is simply 
\begin{equation}
h_{\ell m}^{\Newt,\epsilon_p} \equiv \frac{M_{\MBH}}{R}\, n^{(\epsilon_p)}_{\ell m}\, c_{\ell +\epsilon_p}\, v^{\ell +\epsilon_p}\, 
Y_{\ell - \epsilon_p,-m}(\pi/2,\phi).
\end{equation}
where $Y_{\ell,m}(\theta,\phi)$ are spherical harmonic functions, while $n_{\ell m}^{(\epsilon_p)}$ and $c_{\ell +\epsilon_p}$ are numerical coefficients~\cite{Damour:2008gu}. 

The flux presented above only accounts for the amount of radiation that escapes to infinity ${\cal{F}}^{g}_{\GR,\infty}$, to which one must add the amount of radiation absorbed by the supermassive black hole ${\cal{F}}^{g}_{\GR,+}$. In the EOB-EMRI framework, this is accounted for by linearly adding black hole absorption terms and calibration coefficients, fitted to a more accurate, numerical flux~\cite{Yunes:2010zj}. The calibration accounts for unknown, high-order PN terms. 

With all of this at hand, one can solve the differential system in Eq.~\eqref{omegaadiab} with certain post-circular initial conditions~\cite{Buonanno00}. We implement this by first performing a mock-evolution that starts at $100 M_{\MBH}$ with post-circular initial conditions (see e.g.~\cite{Yunes:2010zj}) and then reading off $[\dot{\phi}(0),\phi(0)]$ at the desired starting point of the real evolution. The gravitational wave phase and amplitude can be obtained from the waveforms via
\begin{equation}
\Phi_{\GW}^{\ell m} = \Im \left[ \ln \left(\frac{h_{\ell m}}{|h_{\ell m}|} \right)\right]\,, \qquad
A_{\GW}^{\ell m} = |h_{\ell m}|\,.
\end{equation}

Equations~\eqref{E-binding} and~\eqref{pd-flux-adiab} neglect the effect of the small compact object on its own evolution, i.e.~the correction to the metric, Hamiltonian and radiation-reaction due to the small compact object's self-gravity. These correspond to conservative and second-order dissipative self-force effects, which we will here neglect. Although such effects are important for the modeling of EMRI waveforms, they are not currently known in the strong-field regime and thus cannot be properly incorporated into this framework (however, see the recent work~\cite{Warburton:2011fk} where such effects have been included in the case of Schwarzschild black holes within GR). 

\subsection{Scalar-Tensor Modification \\ to the EOB-EMRI Model}

The EOB-EMRI framework described in the previous subsection can be modified to model EMRIs in scalar-tensor theories by properly accounting for all energy sinks. In scalar-tensor theories, the gravitational wave energy flux is complemented by the scalar field energy flux: 
\be
{\cal{F}}_{\Tot} = {\cal{F}}^{g}_{\GR,\infty} + {\cal{F}}^{g}_{\GR,+} + {\cal{F}}^{s}_{\infty} + {\cal{F}}^{s}_{+}\,.
\ee
Depending on the particular scalar-tensor theory studied (eg.~massive or massless), ${\cal{F}}^{s}$ may or may not contain resonances. Massless ($\mu_s=0$) scalar-tensor theories are considered in Sec.~\ref{prof-const-massless}, by modeling the energy flux via the numerical results presented in Sec.~\ref{Sec:EfluxMassless}. The detectability of massive scalar-tensor theories is discussed in Sec.~\ref{sec:Constraints-Massive}. Of course, the modification of the energy flux due to scalar emission will induce a non-GR radiation-reaction force that, although small at any particular time (a part from possible resonant effects), will build up possibly leading to strong modifications in the waveforms.

In addition to this, there is in principle a modification to the binding energy. This effect, however, is difficult or impossible to measure with gravitational waves. This is because in Brans-Dicke theory, the scalar field can be thought of as renormalizing Newton's constant, or equivalently the bare system parameters. For example, given a gravitational wave detection, one would think one is measuring the bare component masses, while in reality one would be measuring a certain component tensor mass, composed of the product of the inertial masses and the Brans-Dicke coupling parameter~\cite{Fujii_Maeda_book,lrr-2006-3}. For this reasons, we will ignore these effects in the current paper, and only model dipolar corrections due to scalar field emission. 

\subsection{Projected Constraints on Massless Scalar-Tensor Theories}
\label{prof-const-massless}

Consider the following two systems:
\begin{itemize}
\item {\bf{System~F}}: $M_{\MBH} = 10^{7} \, M_{\odot}$, $m_{\SCO} = 1.4 \, M_{\odot}$ ($\nu=1.4\times10^{-7}$), $q_{\MBH} = 0.99$, $r/M_{\MBH} \in (1.927,1.455)$, $v \in (0.649,0.714)$, $f_{\rm GW} \in (0.00176,0.00235) \; {\rm{Hz}}$, total cycles in $\sim 3.75 \times 10^{5}$ rads of gravitational wave phase.
\item {\bf{System~R}}: $M_{\MBH} = 10^{5} \, M_{\odot}$, $m_{\SCO} = 1.4 \, M_{\odot}$ ($\nu=1.4 \times 10^{-5}$), $q_{\MBH} = 0.7$, $r/M_{\MBH} \in (18.145,15.985)$, $v \in (0.234,0.249)$, $f_{\rm GW} \in (0.008,0.010) \; {\rm{Hz}}$, total cycles in $\sim 1.79 \times 10^{6}$ rads of gravitational wave phase.
\end{itemize}
The one-year evolution of System~F within a LISA-like band samples a strong-field region and it is representative for a configuration which maximizes the relativistic effects of the scalar emission. The small compact object orbits very close to the supermassive black hole and indeed the evolution is stopped at the orbital separation corresponding to the ISCO. On the other hand, System~R is a more realistic configuration, in which the supermassive black hole is rapidly spinning, but is not extremal, while the orbital velocity and the mass ratio are  optimal for the new eLISA configuration~\cite{AmaroSeoane:2012km}. In this case, the EMRI samples a moderately strong-field regime and the evolution is stopped at the orbital radius corresponding to a final gravitational wave frequency of $0.01$Hz.

The calculation of the gravitational wave phase in massless scalar-tensor theories will depend not only on the value of $\alpha$, or equivalently $\omega_{\BD}$, but also on the sensitivity parameter $s_{\SCO}$. As for the former, we will here work directly in terms of $\omega_{\BD}$ and we will saturate the Solar System constraint $\omega_{\BD} = 4 \times 10^{4}$. This will allows us to determine whether such a value of $\omega_{\BD}$ leads to an observable effect in EMRI waveforms. As for $s_{\SCO}$, we use the representative value $s_{\SCO} \approx 0.188$, which corresponds to the average of three sensitivities for three different neutron stars of the same mass but with different equations of state~\cite{1989ApJ...346..366W}.

Before comparing waveforms directly, we will perform certain shifts in time and phase so that the waveforms are initially aligned in phase and frequency. In particular, we minimize the statistic in Eq.~(23) of Ref.~\cite{Buonanno:2009qa}, which is equivalent to maximizing the fitting factor over time and phase of coalescence in a matched filtering calculation with white noise~\cite{Buonanno:2009qa}. We carry this out in the low-frequency regime, inside the time interval $(0,64) \lambda_{\GW}$ (roughly $0.01$ months), where $\lambda_{\GW}$ is the gravitational wave wavelength). 

Let us now compare the phase and the amplitude of GR and massless scalar-tensor theory waveforms. We define the {\emph{dephasing}} by $\delta \phi = \phi_{\GR} - \phi_{\ST}$, measured in radians, where $\phi_{\GR}$ is the waveform phase in GR and $\phi_{\ST}$ is the waveform phase in scalar-tensor theories. 
We define the {\emph{normalized amplitude difference}} by $\delta A = |1 - A_{\ST}/A_{\GR}|$, where $A_{\GR}$ is the GR waveform amplitude and $A_{\ST}$ is the waveform amplitude in scalar-tensor theories. 
\begin{figure*}
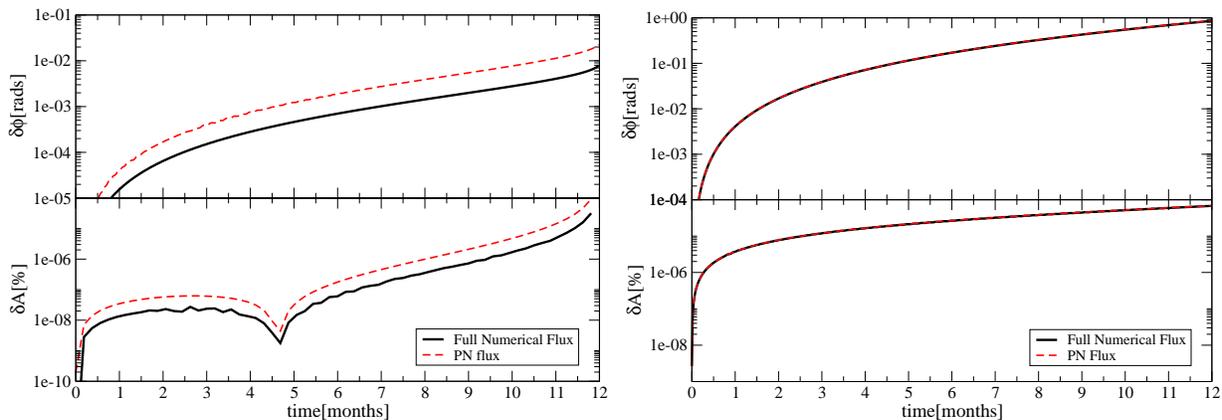

\begin{center}
\begin{tabular}{cc}
 \epsfig{file=Phase-Comp_F.eps,width=8cm,angle=0,clip=true} &
 \epsfig{file=Phase-Comp_R.eps,width=8cm,angle=0,clip=true}
\end{tabular}
 \caption{\label{fig:flux-comp_F} Dephasing and normalized amplitude difference as a function of time in units of months for System~F (left) and System~R (right). The black and red curves are obtained by using the numerical fluxes constructed in Sect.~\ref{sec:GWST} and their PN approximation in Eq.~\eqref{dipole-massless}, respectively. We consider a Brans-Dicke parameter which saturates the Solar System bound, $\omega_{\BD} = 4 \times 10^{4}$.}
\end{center}
\end{figure*}
Figure~\ref{fig:flux-comp_F} shows the dephasing and normalized amplitude difference as a function of time in units of months for Systems F (left) and R (right), respectively. We show results obtained using both our numerical flux and the PN formula in Eq.~\eqref{dipole-massless}.

Our results show that EMRIs cannot place constrains on massless scalar-tensor theories, including Brans-Dicke theory, that are stronger than current Solar System ones. Figure~\ref{fig:flux-comp_F} shows that, even considering one of the most relativistic cases among detectable EMRI systems, the accumulated dephasing over one year is less than $\delta\phi\sim0.1$, which is roughly expected to be the limit for detection with eLISA (unless the signal-to-noise ratio is much larger than ten). Of course, if we had chosen $\omega_{\BD} < 4 \times 10^{4}$, then the dephasing would have been larger, as the emitted flux scales as $\omega_{\BD}^{-1}$, but such values of $\omega_{\BD}$ are already ruled out by Solar System experiments. Furthermore, the PN results obtained from Eq.~\eqref{dipole-massless} overestimate the dephasing and the normalized amplitude difference computed with our full numerical flux. This implies that tests of Brans-Dicke theory with EMRIs will be more difficult than previously expected by extrapolating leading-order, PN results to the strong-field region. 

The right-panel panel of Fig.~\ref{fig:flux-comp_F} shows that our results perfectly agree with the PN expectation when considering System~R. Indeed, System~R is less relativistic than System~F, and thus, for the former the $2.5$PN formula of Eq.~\eqref{dipole-massless} is a very good approximation (cf.~Fig~\ref{fig:scalar_fluxes}). Furthermore, in this case the dephasing after one year evolution is larger than that for System~F. This is not only because System~R has accumulated ten times more phase than System~F in the same observation time (because the radiation-reaction time scale is longer), but also because Brans-Dicke theory leads to larger relative deviations when the EMRI is farther away from the ISCO. As we showed in Fig.~\ref{fig:EdotTot}, the massless scalar and GR energy fluxes are of the same order of magnitude for orbital velocities $v \sim v_{\rm th}$, and thus, the radiation-reaction force is most modified for System~R as $v_{\rm th}<v_{\rm Sys~R} < v_{\rm Sys~F}$. This suggests that to get the strongest constraints, one should consider EMRIs with $v \sim v_{\rm th}$, which translates to separations around $r_{\rm th} \sim v_{\rm th}^{-2} \sim 10^{6} M_{\MBH}$. Unfortunately, the larger the orbital separation, the lower the gravitational wave frequency, and for such large separations, the EMRI emits waves completely outside of the sensitivity band of future detectors.

These results go against the expectation that EMRIs will be the best probes of modified gravity theories. This expectation derives from the observation that EMRIs sample the near-horizon region, and thus, the strong-curvature regime of the underlying gravity theory~\cite{Ryan:1995wh}. However, although Brans-Dicke gravity is among the most popular alternatives to Einstein's theory, it is not, by itself, a strong-curvature modification of GR. As Fig.~\ref{fig:EdotTot} showed, Brans-Dicke theory leads corrections that dominate over GR at low inspiral velocities, not close to the horizon. EMRIs should experience much larger deviations in theories that introduce true, strong-curvature corrections, like e.g. Chern-Simons gravity~\cite{Alexander:2009tp} or alternative theories with generic quadratic curvature corrections~\cite{Yunes:2011we}. These theories are poorly constrained in the weak-field regime~\cite{Yunes:2009hc,AliHaimoud:2011fw}, while EMRI gravitational waves are expected to lead to the strongest constraints~\cite{Stein:2010pn,Yunes:2007ss,Sopuerta:2009iy,Pani:2011xj}.

\section{Projected Constrains on Massive Scalar-Tensor Theories}
\label{sec:Constraints-Massive}

Let us now consider the effect of massive scalar-tensor theories on the gravitational wave phase, and try to estimate whether this is observable with future detectors. As we noted in Sec.~\ref{Sec:EfluxMassive}, the total flux in the massive case is generically smaller than the one in the massless case. Therefore, except for resonances in the scalar flux~\cite{Cardoso:2011xi}, the dephasing induced in the massive case is smaller than in the massless case.  Resonances, on the other hand, can lead to large dephasings, as we will show in this section. 

We begin by considering the region of phase space where resonances are present. We then summarize an analytical prescription to calculate the gravitational wave phase in GR. We choose such an analytical route, instead of an EOB-EMRI one, because the resonances tend to be very difficult to resolve numerically. With this in hand, we then proceed to calculate the accumulated phase when an EMRI traverses a resonance and the corresponding dephasing. 

\subsection{Floating versus Non-Floating Orbits}

\begin{figure}
 \epsfig{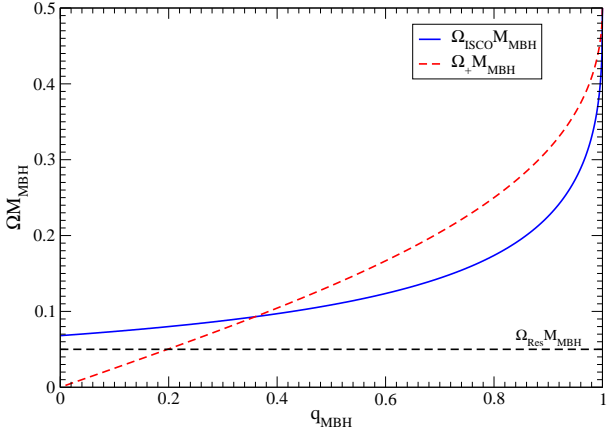}
 \caption{\label{fig:isco} ISCO frequency and angular velocity at the horizon as functions of the spin parameter for a Kerr black hole. In massive scalar tensor theories, resonances occur for particles in circular orbits at $\Omega_\orb\sim\mu_s+{\cal O}( \mu_s^2)$. Stability of the orbit implies $\Omega_\orb<\Omega_\ISCO$, i.e. the relevant region is below the blue line. For $q_\MBH\gtrsim0.36$, these frequencies always satisfy the superradiant condition for $m=1$, $\Omega_\orb<\Omega_+$. For $q_\MBH\lesssim0.36$, the resonant frequency may also be non-superradiant in the region limited by the blue line and the red dashed line. Note that resonances may occur even for Schwarzschild black holes.}
\end{figure}

Depending on the spin of the massive black hole, the resonances in the scalar horizon flux can lead to two qualitatively different effects. As shown in Fig.~\ref{fig:isco}, if $q_\MBH\gtrsim0.36$ any orbital frequency $\Omega_\orb<\Omega_\ISCO(q_\MBH)$ also satisfies the superradiant condition for $m=1$, $\Omega_\orb<\Omega_+(q_\MBH)$. Thus, possible resonant fluxes are \emph{negative} in this region. On the other hand, if  $q_\MBH\lesssim0.36$, non-superradiant frequencies are also excited and the resonant flux becomes \emph{positive}. 

A negative flux leads to the interesting possibility of a {\emph{floating orbit}}, i.e.~an orbit where ${\cal{F}}^{g}_{\GR}+{\cal{F}}^{s,\massive}=0$~\cite{Cardoso:2011xi}. Recall, however, that this does not imply the orbit freezes at a given radius, but rather that its inspiral slows down, as it is not driven by the shedding of mass and spin angular momentum of the supermassive black hole. The floating orbit condition on the fluxes defines the region in $(\mu_{s},\alpha)$ space where floating can occur:
\begin{equation}
 \alpha>\alpha_c=\frac{16 \sqrt{\pi}}{\sqrt{15}}\sqrt{q_\MBH-2y(1+\sqrt{1-q_\MBH^2})}y^{11/6}\,.\label{alpha_crit}
\end{equation}
where we have assumed $y=\mu_{s} M_{\MBH} \ll1$ and recall that, by Eq.~\eqref{EQBD1a}, the leading-order quadrupolar flux reads ${\cal{F}}^{g}_{\GR} = (32/5) \nu^{2}  v^{10}$. Equation~\eqref{alpha_crit} ensures that $k_+<0$ at the {\emph{floating frequency}} $\Omega_\orb=\Omega_{\float}$. Notice that the floating frequency is smaller, but very close to, the resonance frequency, with differences of ${\cal{O}}(\mu_{s}^{9})$. This, however, does not imply that ${\cal{F}}^{s,\massive}_{+}(\Omega_{\res}) = {\cal{F}}^{s,\massive}_{+}(\Omega_{\float})$, since ${\cal{F}}^{s,\massive}_{+}$ is a very steep function of $\Omega$ close to resonance. 

\begin{figure}
 \epsfig{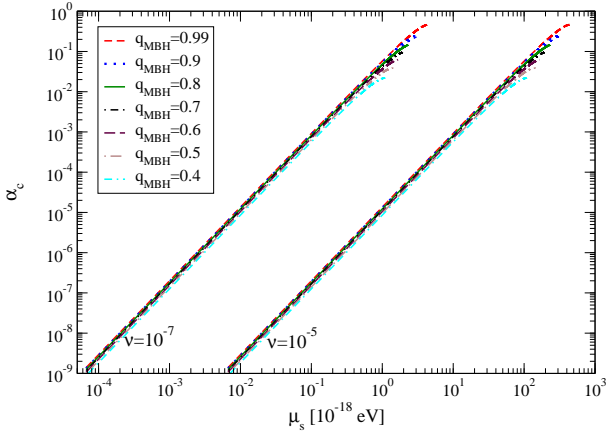}
 \caption{\label{fig:alpha_crit} Critical value, $\alpha_c$, as a function of the scalar mass $\mu_s$, for several values of $q_\MBH$ and for $M_\MBH=10^5 M_\odot$ (lower curves) and $M_\MBH=10^7 M_\odot$ (upper curves). Floating orbit exists for those values of $\alpha$ above the curves and for $q\gtrsim0.36$. Curves terminate when $\mu_s\sim \Omega_\ISCO$.}
\end{figure}

In Fig.~\ref{fig:alpha_crit} we plot $\alpha_c$ in Eq.~\eqref{alpha_crit} as a function of the scalar mass, $\mu_{s}$, in units of $10^{-18}$ eV for two systems where the supermassive black hole has mass $10^{5} M_{\odot}$ and $10^{7} M_{\odot}$ and different spins. Notice that $\alpha_c$ is essentially insensitive to the spin of the supermassive black hole, while for different symmetric mass ratios $\nu$, $\alpha_{c}$ shifts horizontally.

In the particular case of massive Brans-Dicke theory, we can also compute the region of the $(\omega_{\BD},\mu_{s})$ parameter space where floating can occur. Using Eq.~\eqref{alpha_crit} and the definition in Eq.~\eqref{def_alpha} of $\omega_{\BD}$ in terms of $\alpha$, we find
\begin{eqnarray}
 \omega_{\BD}<\omega_{\BD}^{c}&=&-\frac{3}{2}+\frac{15}{16}\left(s_\SCO-\frac{1}{2}\right)^2\times\nn\\
&\times&\frac{\left(\mu_s M_\MBH\right)^{-11/3}}{q_\MBH-2(1+\sqrt{1-q_\MBH^2})\mu_s M_\MBH} \,,\nn\\\label{omegaBD_crit_Einstein}
\end{eqnarray}
The regions in $(\omega_{\BD},\mu_{s})$ phase space where floating occurs is shown in Fig.~\ref{fig:floating-regions} for a supermassive black hole with $M_{\MBH} = 10^{5} M_{\odot}$, spin $q_{\MBH} = 0.9$ and a neutron star sensitivity of $s_{\SCO} = 0.188$. The red-dashed curve is the boundary between floating and non-floating orbits for this system, given by Eq.~\eqref{omegaBD_crit_Einstein}: non-floating resonances exist for values of $(\omega_{\BD},\mu_{s})$ above this line, while floating resonances exist for values below it. For comparison, we plot the $(\omega_{\BD},\mu_{s})$ region (below the solid black line) that is ruled out by the tracking of the Cassini spacecraft~\cite{Perivolaropoulos:2009ak,Alsing:2011er}. Notice that as $\mu_{s} \to 0$, one recovers the massless Brans-Dicke bound $\omega_{\BD} \geq 4 \times 10^{4}$. Since the resonance frequency scales as $\Omega_{\res} \sim \mu_{s}$ (cf.~Eq.~\eqref{dipole}), we also show the values of $\mu_{s}$ (green, dot-dashed curve) for which the resonant frequency occurs inside the classic-LISA sensitivity band [$(10^{-4},10^{-2})$ Hz]. Finally, the vertical dotted line denotes the ISCO frequency, $\Omega_\ISCO$. Curves in Fig.~\eqref{fig:floating-regions} terminate approximately when $\mu_s\sim\Omega_\ISCO$, which corresponds to the small compact object reaching the ISCO.


Just because a floating or non-floating resonance exists does not necessarily mean that it will lead to a detectable effect in the waveform. Whether such resonances will matter or not depends both on the height of the resonance and the amount of time it is active. In the non-floating case, this lifetime is just the width of the resonance $\delta t_{r}$. In the floating case, the resonance width does not come into play, as floating orbits are slowed down and inspiral due to the supermassive black hole's backreaction. In the latter, one must wait a time $\delta \tau$ until the supermassive black hole loses enough mass and spin angular momentum so that the height of the floating resonance is decreased, allowing the small object to pass through. In what follows we will study how both types of resonances affect the waveform and whether they lead to observable effects. 

\subsection{Analytical Modeling of the Gravitational Wave Phase in General Relativity}
\label{analytic-GR-model}

Let us first consider two of the most popular methods to compute the gravitational wave phase in GR. The first method consists of employing the balance law 
\be
\dot{E}_{b}^{\GR} = - {\cal{F}}^{g}_{\GR}\,,
\label{balance-GR}
\ee
where $\dot{E}_{b}$ is the time derivative of the binding energy and ${\cal{F}}^{g}_{\GR}$ is the total flux of energy carried away from the binary system. The binding energy can be computed from the Hamiltonian of the system, and to leading, Newtonian order it is simply 
\be
E_{b}^{\GR} = \nu M_{\MBH} - \frac{\nu M_{\MBH}}{2} \gamma\,,
\label{binding-E-GR}
\ee
where we have defined $\gamma = M_{\MBH}/r$. Through the Hamiltonian, one can also establish Kepler's third law, namely that
\be
\gamma = x \equiv \left(M_{\MBH} \Omega_{\orb}\right)^{2/3}\,,
\ee
which then leads to $\dot E_{b}^\GR = -\dot x (\nu M_{\MBH})/2 $. The total energy flux in GR is simply given by the total gravitational wave energy flux, which to leading-Newtonian order is simply
\be
{\cal{F}}^{g}_{\GR} = \frac{32}{5} \nu^{2} x^{5}\,,
\ee
where again we have used Kepler's law to relate $\gamma$ to $x$. 

With this at hand, we can use the balance equation~\eqref{balance-GR} to derive a differential equation for the frequency evolution, namely
\be
\dot{x}^{\GR} = \frac{64}{5} \frac{\nu}{M_{\MBH}} x^{5}\,,
\ee
or in terms of the orbital frequency $\Omega_{\orb} = 2 \pi F_{\orb}$, 
\be
\dot{F}_{\orb}^{\GR} = \frac{48}{5 \pi} \frac{\nu}{M_{\MBH}^{2}} \left(2 \pi M_{\MBH} F_{\orb}\right)^{11/3}\,.
\ee
These differential equations are separable, and thus, we can solve them to obtain
\begin{align}
x^{\GR}(\Theta) &= \frac{1}{4} \Theta^{-1/4}\,,
\\
F_{\orb}^{\GR}(\Theta) &= \frac{4^{-3/2}}{2 \pi M_{\MBH}} \Theta^{-3/8}\,,
\label{GR-sol-F}
\end{align}
where we have defined the quantity 
\be
\Theta(t) \equiv \frac{\nu}{5 M_{\MBH}} \left(t_{c} - t\right)\,.
\ee
The quantity $t_{c}$ is a constant of integration chosen such that $t = t_{c}$ when $F_{\orb} = +\infty$. This constant is sometimes referred to as the time of coalescence, and thus, $t_{c} - t > 0$. 

With such a frequency evolution, it is simple to obtain the GW phase. For the dominant harmonic mode in a quasi-circular inspiral, this phase is related to the orbital one via $\phi_{\GW} = 2\phi_{\orb}$, and thus
\be
\phi_{\GW}^{\GR} = 4 \pi \int^{t} F_{\orb}(t') \; dt' = -\frac{10}{\nu} \int^{\Theta} x^{3/2}(\Theta') \; d\Theta'\,.
\label{GW-phase}
\ee
We can solve this integral to find
\be
\phi_{\GW}^{\GR}(\Theta) = - \frac{x^{-5/2}}{16\nu} = - \frac{2}{\nu} \Theta^{5/8}\,.
\label{GR-GW-phase}
\ee
This expression agrees identically, to leading, Newtonian order, with PN results~\cite{Blanchet:2002av}. Inserting typical values, we find
\begin{align}
\left|\phi_{\GW}^{\GR}(t)\right| &=  \frac{1}{\nu} \left[\frac{\nu}{5 M_{\MBH}} \left(t_{c} - t\right)\right]^{5/8}\,.
\nn \\
&= 4 \times 10^{6} {\rm{rads}} 
\left(\frac{10^{-5}}{\nu}\right)^{3/8} 
\nn \\
&\times \left(\frac{10^{5} M_{\odot}}{M_{\MBH}}\right)^{5/8} 
\left(\frac{t_{c} - t}{1 {\rm{yr}}}\right)^{5/8}\!\!\!\!\!\!,
\end{align}

The GW phase in Eq.~\eqref{GR-GW-phase} is exact to leading, Newtonian order, but it might not be very useful for EMRIs, since it assumes that the orbiting body reaches merger in a one year evolution. EMRIs, however, can be in orbit for tens of years if their mass ratio is small enough. When dealing with EMRIs that might not merge in a one-year evolution, it is perhaps more appropriate to rederive the GW phase in a slightly different way. Consider then the following expression for the GW phase
\be
\phi_{\GW}^{\GR} = 2 \int_{r_{\initial}}^{r_{\final}} \Omega_{\orb}(r) \; \frac{dr}{\dot{r}^{\GR}}\,,
\label{GR-GW-phase-for-EMRIs-integral}
\ee
where $(r_{\initial},r_{\final})$ are the initial and final orbital radius that are observed, while $\Omega_{\orb} = \gamma^{3/2}/M_{\MBH}$ by Kepler's third law, with the rate of radial inspiral $\dot{r}^{\GR}$ obtained by the chain rule:
\be
\dot{r}^{\GR} = \frac{dE_{b}^{\GR}}{dt} \frac{dr}{dE_{b}^{\GR}} = -{\cal{F}}^{g}_{\GR} \left(\frac{dE_{b}^{\GR}}{dr}\right)^{-1} = - \frac{64}{5} \nu \gamma^{3}\,.
\label{radial-evol-eq-GR}
\ee
The GW phase in GR is then simply
\be
\phi_{\GW}^{\GR} = \frac{1}{16 \nu} 
\left[\left(\frac{r_{\initial}}{M_{\MBH}}\right)^{5/2} - \left(\frac{r_{\final}}{M_{\MBH}}\right)^{5/2} \right]\,. 
\label{GR-GW-phase-for-EMRIs}
\ee

The GW expression in Eq.~\eqref{GR-GW-phase-for-EMRIs} can be converted into a function of time by solving the radial evolution equation. The solution to Eq.~\eqref{radial-evol-eq-GR} is simply
\be
T_{\final,\initial}^{\GR} = \frac{5}{256} \frac{1}{\nu} \left[ \left(\frac{r_{\final}}{M_{\MBH}}\right)^{4} - \left(\frac{r_{\initial}}{M_{\MBH}}\right)^{4} \right]\,,
\label{Tfi-rel}
\ee
where $T_{\final,\initial}$ is the time for the small object to go from $r_{\initial}$ to $r_{\final}$. Using this in Eq.~\eqref{GR-GW-phase-for-EMRIs}, we obtain
\begin{align}
\phi_{\GW}^{\GR} &= \frac{1}{16 \nu} \left(\frac{r_{\final}}{M_{\MBH}}\right)^{5/2}
\left[\left(\frac{r_{\initial}}{r_{\final}}\right)^{5/2} - 1 \right]\,, 
\nn \\
 &= \frac{1}{16\nu}  \left(\frac{r_{\final}}{M_{\MBH}}\right)^{5/2}
\nn \\
&\times
\left\{ \left[ 1 + \frac{256}{5} \nu \frac{T_{\final,\initial}^{\GR}}{M_{\MBH}} \left(\frac{M_{\MBH}}{r_{\final}}\right)^{4} \right]^{5/8} - 1 \right\}\,.
\label{GR-GW-phase-for-EMRIs-2}
\end{align}
The time $T_{\final,\initial}^{\GR}$ can be required to equal the observation time, i.e.~$1$ year. This expression agrees with that used, e.g.~in~\cite{Yunes:2011ws,Kocsis:2011dr}, and it reduces to Eq.~\eqref{GR-GW-phase-for-EMRIs} when the second term in square brackets dominates. 

\subsection{Effect of Non-Floating Resonances \\ on the Gravitational Wave Phase}
\label{sec:resonant-effect-on-GW}

Let us now concentrate on the gravitational wave phase as an EMRI crosses a non-floating resonances. We model the correction to the GR gravitational wave flux via ${\cal{F}} = {\cal{F}}^{g}_{\GR} + {\cal{F}}^{s}$, with 
\be
{\cal{F}}^{s}(t) = h(t) \; {\cal H}\left(\delta t_{\res}^{2} - \left(t - t_{\res}\right)^{2}\right)\,,
\label{non-float-res-model}
\ee
where $h(t)$ is the height of the resonance, $t_{\res}$ is the time at which the resonance occurs, $\delta t_{\res}$ is its width, where recall that ${\cal H}(x)$ is the Heaviside function. Notice that Eq.~\eqref{non-float-res-model} is nothing but a top-hat function of width $\delta t_{\res}$, centered at $t = t_{\res}$ and with a time-dependent height $h(t)$. The height $h(t)$ represents the total scalar flux ${\cal{F}}_{\Tot}^{s,\massive}$, which must satisfy $|h(t)| < |{\cal{F}}^{g}_{\GR}(t)|$ for non-floating resonances. If $h(t) > 0$, one has a {\emph{sinking resonance}}, i.e.~a resonance whose modification to the flux enhances the inspiral rate. Of course, one can also have $h(t)<0$ but still $|h(t)| < |{\cal{F}}^{g}_{\GR}|$ such that the orbit is non-floating and non-sinking.  

For massive scalar-tensor theories, we can compute $\delta t_{\res}$ explicitly. This width can be obtained from
\be
 \delta t_{\res} \equiv \frac{|\Delta \Omega|}{\dot \Omega^{\res}_{\orb}}\,,
\ee
where $\Delta \Omega$ is the resonance width in frequency space [cf.~Eq.~\eqref{width}]. We can model $\dot{\Omega}_{\orb}$ at resonance as 
\begin{align}
\dot{\Omega}_{\orb}^{\res} &= \left[\frac{d{E}_{b}}{dt} \left(\frac{dE_{b}}{d\Omega}\right)^{-1}\right]_{\res}\,,
\nn \\
&= - \left({\cal{F}}^{g}_{\GR} + {\cal{F}}^{s} \right)_{\res} \left(\frac{dE_{b}}{d\Omega}\right)^{-1}_{\res}\,, 
\nn \\
&\sim - \left[{\cal{F}}^{g}_{\GR} \left(\frac{dE_{b}}{d\Omega}\right)^{-1}\right]_{\res}\,, 
\end{align}
where in the last line we have neglected the scalar flux, assuming that in general this is smaller than the gravitational one. This then leads to
\begin{align}
 \delta t_{\res}  &=\frac{5}{96} \frac{M_{\MBH} }{\nu} \left(M_{\MBH} \Omega_{\res}\right)^{-11/3} \left(M_{\MBH} \Delta \Omega\right)\,,
\end{align}
where we have used that ${\cal{F}}^{g}_{\GR} = (32/5) \nu^{2}  x^{5}$. Substituting in $\Delta \Omega = \Delta \Omega^{lmn}$ from Eq.~\eqref{width-simplified} and $\Omega_{\res}$ from Eq.~\eqref{om_resonance}, we find
\begin{align}
 \delta t_{\res} &= \frac{5}{1152} \frac{M_{\MBH}}{\nu} \left(\mu_{s} M_{\MBH}\right)^{16/3} \left(q_{\MBH} - 2 r_{+} \mu_{s}\right)\,,
 \nonumber \\
 &= 10^{-16} \; {\rm{yrs}} \; \left(\frac{10^{-5}}{\nu}\right) \left(\frac{M_{\MBH}}{10^{5} M_{\odot}}\right)  
  \left(\frac{\mu_{s} M_{\MBH}}{0.01}\right)^{16/3} 
  \nonumber \\
  &\times
  \left(\frac{q_{\MBH} - 2 r_{+} \mu_{s}}{0.9 - 2 \times 1.43 M_{\MBH} \times 0.01 M_{\MBH}^{-1}}\right)\,,
  \label{widthevaled}
\end{align}
where in the second equality we have scaled the width by typical numbers. Clearly then, $\delta t_{\res} \ll T_{\obs}$ and we can treat $h(t) = h$ as a constant, i.e.~ $h= {\cal{F}}_{+}^{s,\rm{massive}}$ in Eq.~\eqref{res-height} with $\alpha < \alpha_{c}$. 

Let us now use the methods in Sec.~\ref{analytic-GR-model} to derive the gravitational wave phase as an EMRI traverses a non-floating resonance. Let us then use the balance equation to derive a differential equation for the orbital frequency evolution:
\be
\dot{E}_{b} = -{\cal{F}}^{g}_{\GR} - {\cal{F}}^{s}\,,
\ee
where ${\cal{F}}^{g}_{\GR}$ is the GR gravitational wave energy flux and ${\cal{F}}^{s}$ is the scalar energy flux of Eq.~\eqref{non-float-res-model}. Since the Hamiltonian of the system is not modified by the scalar resonance, we still have that $E_{b}$ is given by Eq.~\eqref{binding-E-GR} to leading-order and Kepler's third law still holds. 

The evolution equation for the orbital frequency is then
\begin{align}
\dot{F}_{\orb} &= \frac{48}{5 \pi} \frac{\nu}{M_{\MBH}^{2}} \left(2 \pi M_{\MBH} F_{\orb}\right)^{11/3} 
\nn \\
&+ \frac{3}{2 \pi} \frac{1}{ \nu M_{\MBH}^{2}} \left(2 \pi M_\MBH F_{\orb}\right)^{1/3} {\cal{F}}^{s}.
\label{non-float-DE-for-F}
\end{align}
This is a differential equation for $F_{\orb}$ that is not separable due to the ${\cal{F}}^{s}$ term. However, since we are dealing with non-floating resonances, we can search for solutions $F_{\orb} = F_{\orb}^{\GR} + \delta F_{\orb}$, where $\delta F_{\orb}$ is assumed to be smaller than $F_{\orb}^{\GR}$. Inserting this ansatz into Eq.~\eqref{non-float-DE-for-F} we obtain the zeroth-order differential equation
\be
\dot{F}_{\orb}^{\GR} = \frac{48}{5 \pi} \frac{\nu}{M_{\MBH}^{2}} \left(2 \pi M_{\MBH} F_{\orb}^{\GR}\right)^{11/3}\,,
\ee
and the first-order differential equation
\ba
\delta \dot{F}_{\orb} &=& \frac{352}{5} \nu  \left(2 \pi M_{\MBH} F_{\orb}^{\GR}\right)^{8/3} \frac{\delta F_{\orb}}{M_{\MBH}}  
\nonumber \\
&+& \frac{3}{2 \pi} \frac{1}{\nu M_{\MBH}^{2}} \left(2 \pi M_{\MBH} F_{\orb}^{\GR} \right)^{1/3} {\cal{F}}^{s}\,.
\ea

The solution to these differential equations is simple. The solution to the zeroth-order equation is that of Eq.~\eqref{GR-sol-F}. The solution for the correction to the frequency evolution can be written as
\be
\delta F_{\orb} = \frac{3}{2 \pi} \frac{1}{\nu M_{\MBH}^{2}} e^{\chi} \int^{t} \left(2 \pi M_{\MBH} F_{\orb}^{\GR}\right)^{1/3} e^{-\chi} {\cal{F}}^{s} dt'\,,
\label{delta-F-1}
\ee
where we have set the integration constant to zero and we have defined the quantity
\begin{align}
\chi &\equiv \frac{352}{5} \frac{\nu}{M_{\MBH}} \int^{t} \left(2 \pi M_{\MBH} F_{\orb}^{\GR}\right)^{8/3} dt'\,,
\nonumber \\
&= \frac{11}{2} \int^{x_{\GR}} \frac{dx'}{x'} = \ln\left(x_{\GR}^{11/2}\right)\,, 
\end{align}
With this at hand, Eq.~\eqref{delta-F-1} becomes
\begin{align}
\delta F_{\orb} &= \frac{3}{2 \pi} \frac{x_{\GR}^{11/2}}{\nu M_{\MBH}^{2}}  \int^{t} x_{\GR}^{-5}(t') \; {\cal{F}}_{s}(t') \; dt'\,,
\nonumber \\ 
&= \frac{3}{2 \pi} \frac{x_{\GR}^{1/2}(t_{\res})}{\nu M_{\MBH}^{2}} h \; \delta t_{\res}\,,
\end{align}
where in the second line we have performed the integral over the top-hat function ${\cal{F}}_{s}$ and evaluated all quantities at $t = t_{\res}$, as this is the only time the resonance is active.

From this expression, we can now find the waveform phase. Using Eq.~\eqref{GW-phase}, the gravitational wave phase is
\be
\phi_{\GW}
\sim \phi_{\GW}^{\GR}   + 6 {\cal{M}}^{-2} u_{\res}^{1/3} \; h \; \delta t_{\res}\; T_{\obs}\,,
\ee
where we have used that $u(f) = \pi {\cal{M}} f$, with $f$ the gravitational wave frequency, and the second term integrates simply to $T_{\rm obs}$ because it is constant. The gravitational wave dephasing is then simply
\be
\left|\delta \phi_{\GW}\right| = 6 \; {\cal{M}}^{-2} u_{\res}^{1/3} \; h \; \delta t_{\res} \; T_{\obs}\,.
\label{deph-resonance}
\ee
Of course, recall that this expression is only valid when ${\cal{F}}^{s}$ is assumed much smaller in magnitude than ${\cal{F}}^{g}_{\GR}$. 

The above results are generically applicable to any non-floating resonance that is active for a very short time, such that their modifications to the flux can be treated as effectively a constant small relative to the GR flux. In particular, the above dephasing is also applicable to the GR resonances discovered by Flanagan and Hinderer~\cite{2010arXiv1009.4923F,Gair:2011mr}. Requiring that the resonances lead to an observable effect (i.e.~a dephasing larger than 1 radian), the product of the timescale and the magnitude of the flux relative to the GR flux must satisfy 
\begin{align}
\left(\frac{h}{{\cal{F}}^{g}_{\GR}}\right)\left(\frac{\delta t_{\res}}{M_{\MBH}}\right)  
&\gtrsim \frac{1}{6}  \frac{{\cal{M}}^{2} \; u_{\res}^{-1/3}}{T_{\obs} M_{\MBH} {\cal{F}}^{g}_{\GR}}\,,  
\nonumber \\
&= \frac{5}{196}  \frac{{\cal{M}}}{T_{\obs}} \nu^{3/5} u_{\res}^{-11/3}\; 
\nonumber \\
&\gtrsim 51 \left(\frac{{\cal{M}}}{10^{2} M_{\odot}}\right) 
\left(\frac{1 \; {\rm{yr}}}{T_{\obs}}\right)  
\nonumber \\
&\times
\left(\frac{\nu}{10^{-5}}\right)^{3/5} \left(\frac{u_{\res}}{3 \times 10^{-5}}\right)^{-11/3}\,.
\label{constraint_res}
\end{align}
Equation~\eqref{constraint_res} is an interesting constraint: it implies that if the relative height is of ${\cal{O}}(10^{-1})$, then the resonance width must be {\emph{at least}} $4.2$ seconds wide for the corrections to the GR flux to have an observable effect on EMRI waveforms.

Let us now evaluate this dephasing for the height $h$ and width $\delta t_{\res}$ appropriate to massive scalar tensor theories. Putting all pieces together, the dephasing becomes
\be
\left|\delta \phi_{\GW}\right| = \frac{5}{16} u_{\res}^{-10/3} \; {\cal{F}}^{s,\massive}_{+}(\Omega_{\res}) \; |\Delta \Omega^{lmn}| \; T_{\obs}\,.
\ee
Evaluating this equation to leading-order in $\mu_{s} M_{\MBH} \ll 1$, we obtain
\ba
 |\delta \phi_{\GW}| &\sim& 4.3 \times 10^{-4} \; {\rm{rads}} \; \left(\frac{\alpha}{10}\right)^2 \left(\frac{T_{\obs}}{1 \text{yr}} \right)
\nonumber \\
&\times& 
\left(\frac{10^5 M_{\odot}}{M_{\MBH}}\right) \left[\frac{\mu_s M_\MBH}{0.01}\right]^{16/3}\,,
\ea
where we have rescaled all quantities by typical values. The incredible smallness of this dephasing can be traced back to the fact that the resonance is extremely narrow [cf.~Eq.~\eqref{widthevaled}]. 

We can now require that this dephasing be greater than one radian and solve for the value of $\alpha$ that would allow for this. 
In order to have a dephasing larger than one radian, $\alpha>\alpha_{\rm nf}$, where the latter is defined as
\begin{equation}
 \alpha_{\rm nf}\approx482.9\sqrt{\frac{1 {\rm yr}}{T_\obs}}\sqrt{\frac{M}{10^5 M_\odot}}\left[\frac{\mu_s M_\MBH}{0.01}\right]^{-8/3}\,.
\end{equation}
Alternatively, we can use Eq.~\eqref{def_alpha} and write the corresponding bound for $\omega_{\BD}< \omega_{\BD}^{\rm nf}$, where 
\ba
\omega_{\BD}^{\rm nf} &\approx&  10^{-5}  \left(\frac{\frac{1}{2}-s_\SCO}{\frac{1}{2} - 0.188}\right)^2 \left(\frac{T_{\obs}}{1 \text{yr}} \right)
\nonumber \\
&\times&
\left(\frac{10^5 M_{\odot}}{M_{\MBH}}\right) \left[\frac{\mu_s M_\MBH}{0.01}\right]^{16/3}\,.\label{omega_crit_nf}
\ea
Such a critical value of $\omega_{\BD}$ is shown in Fig.~\ref{fig:floating-regions} as a dot-dashed, blue curve. That is, for values of $(\omega_{\BD},\mu_{s})$ below this curve, the dephasing is larger than one radian, while for values above the curve the dephasing is less than one radian. Observe that the red-dashed curve is always above the blue dot-dashed curve, i.e.~there are no values of $(\omega_{\BD},\mu_{s})$ that can {\emph{simultaneously}} lead to a significant dephasing (larger than 1 rad) and to a non-floating resonance inside the classic-LISA frequency band (green, dot-dashed curve). Mathematically, there are no values of $(\alpha,\mu_{s})$ that can simultaneously satisfy $\alpha > \alpha_{\rm nf}$ and $\alpha < \alpha_{c}$.  
This then automatically implies that non-floating resonances cannot be observed with a LISA-like mission.

\subsection{Effect of Floating Resonances \\ on the Gravitational Wave Phase}
\label{sec:float-cons}

Consider now the gravitational wave phase as an EMRI traverses a floating resonance. During such resonances the total energy flux can be written as
\be
{\cal{F}}_{\Tot} = {\cal{F}}^{g}_{\GR} + {\cal{F}}^{s}\,,
\label{total-floating-flux}
\ee
where we recall that ${\cal{F}}^{g}_{\GR} = {\cal{O}}(\nu^{2})$ is the GR gravitational wave energy flux and ${\cal{F}}^{s}={\cal{O}}(\nu^{2})$ is the scalar energy flux. As we shall see next, if these two terms cancel each other, for example at a floating resonance, an additional {\emph{non-adiabatic}} contribution, ${\cal{F}}_{\rm non-adiab} = {\cal{O}}(\nu^{3})$, arises due to mass and spin angular momentum loss of the supermassive black hole. Therefore, at resonance the binary system continues to lose energy and inspiral, albeit at a slower rate, as the supermassive black hole back-reacts by shedding mass and angular momentum. 

We can make this clearer by considering the balance law $\dot{E}_{b} = - {\cal{F}}_{\Tot}$. The binding energy of the small compact object, Eq.~\eqref{E-binding}, can be written in terms of the orbital frequency as 
\begin{align}
 E_b &=\frac{M_\MBH \nu}{\sqrt{1-v^3 q_\MBH  } }
\frac{1-2 v^2 (1-v^3 q_\MBH  )^{1/3}  }{\sqrt{1+v^3 q_\MBH  -3 v^2 (1-v^3 q_\MBH  )^{1/3}}}\,,
\label{full-GR-binding-E}
\end{align}
where we recall that $v=(M_\MBH\Omega_\orb)^{1/3}$. Equation~\eqref{full-GR-binding-E} asymptotes to $E_{b} \sim \nu M_{\MBH} -(\nu M_{\MBH}/2) v^{2}$ in the $v \ll 1$ limit, which agrees with Eq.~\eqref{binding-E-GR}. To leading order in $\nu$, $M_{\MBH}$ and $q_{\MBH}$ are constant, but to next order they evolve:  
\begin{eqnarray}
 \dot{E}_b=\frac{\partial E_b}{\partial M_\MBH}\dot M_\MBH+\frac{\partial E_b}{\partial q_\MBH}\dot q_\MBH+\frac{\partial E_b}{\partial \Omega_\orb}\dot \Omega_\orb\,,\nn\\ \label{dE}
\end{eqnarray}
where a dot denotes a total derivative. We can then use the balance law to solve for $\dot{\Omega}_{\orb}$, namely
\begin{align}
 \left(\frac{\partial E_{b}}{\partial \Omega_{\orb}}\right) \dot{\Omega}_{\orb} &= -{\cal{F}}_{\Tot} 
- \left(\frac{\partial E_b}{\partial M_\MBH}\dot M_\MBH+\frac{\partial E_b}{\partial q_\MBH}\dot q_\MBH \right)\,,
\end{align}
which is exact to all orders in $\nu$. While the first term on the right-hand side is of ${\cal{O}}(\nu^{2})$, the second term (in parenthesis) is of ${\cal{O}}(\nu^{3})$, because $(\dot{M}_{\MBH},\dot{q}_{\MBH})={\cal{O}}(\nu^{2})$ and $(\partial E_{b}/\partial M_{\MBH},\partial E_{b}/\partial q_{\MBH}) = {\cal{O}}(\nu)$. Usually, one thinks of $(M_{\MBH},q_{\MBH})$ as constant, and through the chain rule identifies the left-hand side as simply $\dot{E}_{b}$. In that case, one can effectively think of the term in parenthesis on the right-hand side as the additional ${\cal{F}}_{\nonadiab}$ in Eq.~\eqref{total-floating-flux}. We emphasize that we are not adding an ${\cal{F}}_{\nonadiab}$ contribution to ${\cal{F}}_{\Tot}$, but rather the time-dependence of $(M_{\MBH},q_{\MBH})$ serves as an {\emph{effective}} modification to ${\cal{F}}_{\Tot}$. 

The evolution of the orbital frequency at resonance can be computed explicitly by using the appropriate rates of change of mass and spin angular momentum. First, we recognize that $q_\MBH=J_\MBH/M_\MBH^2$, due to the laws of black hole mechanics. Moreover, for circular orbits we also have that $\delta M_\MBH=\Omega_\orb \delta J_\MBH$, and thus
\begin{equation}
 \dot q_\MBH= \frac{1-2 q_\MBH M_\MBH  \Omega_\orb }{M_\MBH \Omega_\orb }\frac{\dot M_\MBH}{M_\MBH}\,.\label{dq}
\end{equation}
Then, substituting 
\begin{equation}
 \dot M_\MBH(F^{\float}_{\orb})={\cal{F}}^{s}_{+}(F^{\float}_{\orb})\approx -{\cal{F}}^{g}_{\GR} (F^{\float}_{\orb}) \,,\label{dM}
\end{equation}
and Eq.~\eqref{dq} into $\dot{E}_{b} = 0$ with Eq.~\eqref{dE}, and solving for $\dot\Omega_\orb$ we obtain~\cite{Cardoso:2011xi}, 
\begin{equation}
\dot\Omega_{\orb} \sim 32\nu^2 \; M_\MBH^{7/3} \; \Omega^{13/3}\,,\label{dotOmega_f}
\end{equation}
in the small $\mu_s$ limit. Therefore, once an EMRI hits a floating radius, its inspiral evolution slows down, as described by Eq.~\eqref{dotOmega_f}, which is ${\cal{O}}(\nu)$ smaller than the GR evolution of the orbital frequency. This evolution equation kicks in after the particle has hit a floating resonance and it lasts for a time $\delta \tau$. 

The floating resonance lifetime $\delta \tau$ is defined as the time that it takes the supermassive black hole to lose enough mass and angular momentum for the scalar flux to not be able to compensate the gravitational flux, and thus, $\delta \tau$ is controlled by ${\cal{F}}_{\nonadiab}$. The floating lifetime can be computed from 
\begin{equation}
 \delta \tau\equiv \dot E^{s,\massive}_{+,{\rm peak}}
 \left(\frac{\delta\dot E^{s,\massive}_{+,{\rm peak}}}{\delta t}\right)^{-1}\,.
 \label{tau_f_def}
\end{equation}
The rate of change in the resonant peak can be computed by differentiating Eq.~\eqref{res-height} (considered as a function of $\Omega_\orb$, $M_\MBH$ and $q_\MBH$), applying the chain rule as in Eq.~\eqref{dE} and then using the rates of change $\dot \Omega_\orb$ $\dot M_\MBH$ and $\dot q_\MBH$ computed above.  Finally, in the small $\mu_s$ limit, Eq.~\eqref{tau_f_def} reads~\cite{Cardoso:2011xi} 
\ba
\delta \tau &\sim& \frac{5}{32} q_{\MBH} \; {\cal{M}} \; \nu^{-6/5} \; u(f_{\float})^{-7/3}\,,
\nonumber \\
&=& \frac{5}{32} \frac{q_{\MBH}}{\nu^{2}} M_{\MBH} \left(\mu_{s} M_{\MBH}\right)^{-7/3}\,.
\label{timescale_floating}
\ea
where in the first line $u(f_{\float})$ is the reduced floating frequency and in the second line we have used that $\Omega_{\float} \sim \Omega_{\res} \sim \mu_{s}$. 

Rescaling the lifetime by typical numbers, we find
\ba
\delta \tau &\sim& 10^{6} \; {\rm{yrs}} \; \left(\frac{q_{\MBH}}{0.9} \right) \left(\frac{10^{-5}}{\nu}\right)^{2} 
\nonumber \\
&\times&
\left(\frac{M_{\MBH}}{10^{5} M_{\odot}}\right) \left(\frac{\mu_{s} M_{\MBH}}{0.01}\right)^{-7/3}\,,
\label{timescale_floating_evaled}
\ea
Notice that, in this case, $\delta \tau \gg T_{\obs}$, and thus, $h$ cannot be treated as a constant. Notice also that $\delta \tau$ scales with the $-7/3$ power of the floating frequency. This means that for even lower floating frequencies, the timescale can easily exceed the Hubble time. We find that $\delta \tau \geq H_{0}^{-1}$, where $H_{0}$ is the value of the Hubble expansion parameter today, when $\Omega_{\res} \lesssim 1.7 \times 10^{-4} M_{\MBH}^{-1}$, or equivalently $f_{\GW}^{\res} \sim 10^{-4} \; {\rm{Hz}} \; (10^{5} M_{\odot}/M_{\MBH})$, which coincides with the low-frequency edge of the LISA sensitivity band. 

Let us now calculate the gravitational wave phase evolution as an EMRI traverses a floating resonance. We apply the second algorithm of Sec.~\ref{analytic-GR-model} that is better suited to EMRIs, namely that that is derived from Eq.~\eqref{GR-GW-phase-for-EMRIs-integral}:
\be
\phi_{\GW} = 2 \int_{r_{\initial}}^{r_{\final}} \Omega_{\orb}(r) \; \frac{dr}{\dot{r}}\,.
\label{floating-GW-int}
\ee
The quantity $\dot{r}$ can be decomposed into two pieces: (i) one that is valid between the initial radius $r_{\initial}$ and the radius at which the EMRI begins to float $r_{\float}$; (ii) and one that is valid between the floating radius and the final radius $r_{\final}$. When there is no floating, $\dot{r} = \dot{r}_{\GR}$, where the latter is given in Eq.~\eqref{radial-evol-eq-GR}. When there is floating, $\dot{r}$ can be computed considering $r=r(M_\MBH,q_\MBH,\Omega_\orb)$ in Eq.~\eqref{r-of-w} and applying the same chain rule as in Eq.~\eqref{dE}, namely
\begin{eqnarray}
 \dot r_{\float}&=&\frac{\partial r}{\partial M_\MBH}\dot M_\MBH+\frac{\partial r}{\partial q_\MBH}\dot q_\MBH+\frac{\partial r}{\partial \Omega_\orb}\dot \Omega_\orb \nn\\
&=& - \frac{96}{5} \nu^{2} \gamma^{4}\,. \label{dr_chain}
\end{eqnarray}
where again a dot denotes a total derivative and in the last line we have used Eqs.~\eqref{dq}, \eqref{dM} and~\eqref{dotOmega_f}.

The integral in Eq.~\eqref{floating-GW-int} can then be solved to obtain
\begin{align}
\phi_{\GW} &= \frac{1}{16 \nu} \left[ \left(\frac{r_{\float}}{M_{\MBH}}\right)^{5/2} - \left(\frac{r_{\initial}}{M_{\MBH}}\right)^{5/2} \right]
\nn \\
&+ \frac{5}{168 \nu^{2}} \left[ \left(\frac{r_{\float}}{M_{\MBH}}\right)^{7/2} - \left(\frac{r_{\final}}{M_{\MBH}}\right)^{7/2} \right]\,.  
\label{GW-floating-pre-final}
\end{align}

The floating radius, the initial radius and the final radius can all be written in terms of the time it takes the EMRI to traverse a certain distance. When there is no floating, we have the standard relation of Eq.~\eqref{Tfi-rel}, which we can rewrite as
\be
\frac{r_{\initial}}{r_{\float}} = \left[1 + \frac{256}{5} \nu \frac{T_{\initial,\float}}{M_{\MBH}} \left(\frac{M_{\MBH}}{r_{\float}}\right)^{4} \right]^{1/4}\,,
\ee
where $T_{\initial,\float}$ is the time it takes the EMRI to go from $r_{\initial}$ to $r_{\float}$. When there is floating, the relation between radius and time changes because $\dot{r}$ changes:
\be
\frac{r_{\final}}{r_{\float}} = \left[1 - 96 \nu^{2} \frac{T_{\float,\final}}{M_{\MBH}} \left(\frac{M}{r_{\float}}\right)^{5}\right]^{1/5}\,,
\ee
where $T_{\float,\final}$ is the time it takes the EMRI to go from $r_{\float}$ to $r_{\final}$. Noting that $r_{\float}/M_{\MBH} = (M_{\MBH} \Omega_{\float})^{-2/3} = (M_{\MBH} \mu_{s})^{-2/3}$, we can rewrite the above expression as
\be
T_{\float,\final} = \frac{1}{96} \frac{M_{\MBH}}{\nu^{2}}\left[ \left(M_{\MBH} \mu_{s}\right)^{-10/3} - \left(\frac{r_{\final}}{M_{\MBH}}\right)^{5} \right]\,. 
\label{Tfloat}
\ee

The floating time provides information about the values of $\mu_{s}$ for which our treatment here is valid. Since the floating time cannot exceed the observation time, and we are here setting the latter to one year, then 
\begin{align}
M_{\MBH} \mu_{s}  &< \left[96 \frac{\nu^{2}}{M_{\MBH}} T_{\obs} + \left(\frac{r_{\final}}{M_{\MBH}}\right)^{5}\right]^{-3/10}\,.  
\nn \\
&\lesssim  \left(\frac{M_{\MBH}}{r_{\final}}\right)^{3/2} 
\nn \\
&\sim 0.015 \left(\frac{M_{\MBH}}{10^{5} M_{\odot}}\right)^{3/2} \left(\frac{15 M_{\MBH}}{r_{\final}}\right)^{3/2}\,,
\end{align}
where in the third line we rescaled by typical values. Larger values of $\mu_{s}$ would lead to floating that would exceed one-year. Notice that the bound above is perfectly consistent with our initial assumption of $\mu_s M_\MBH\ll1$. On the other hand, for an EMRI observation to present floating resonance modifications, the EMRI had to evolve through frequencies inside which there is a floating resonance, namely
\begin{align}
\pi M_{\MBH} f^{\GW}_{\rm low} &< M_{\MBH} \mu_{s} < \pi M_{\MBH} f_{\rm high}^{\GW}\,,
\end{align}
which implies
\begin{align}
\mu_{s} M_{\MBH} &> 0.0124 \left(\frac{M_{\MBH}}{10^{5} M_{\odot}}\right) \left(\frac{f_{\rm low}^{\GW}}{0.008 \; {\rm{Hz}}} \right)\,,
\\
\mu_{s} M_{\MBH} &< 0.0155 \left(\frac{M_{\MBH}}{10^{5} M_{\odot}}\right) \left(\frac{f_{\rm low}^{\GW}}{0.01 \; {\rm{Hz}}} \right)\,,
\end{align}
where $(f_{\rm low}^{\GW},f_{\rm high}^{\GW})$ are the lowest and highest gravitational wave frequencies of the observed EMRI. We see then that there exists a region in $\mu_{s}$ space for which a resonance would fall inside of the gravitational wave frequencies sampled by the EMRI, while at the same time leading to floating for less than one year. 

With this at hand, we can rewrite the gravitational wave phase of Eq.~\eqref{GW-floating-pre-final} in a simpler form. To do so, we note that $T_{\obs} = T_{\initial,\float} + T_{\float,\final}$, so that then
\begin{widetext}
\begin{align}
\phi_{\GW} &= \frac{1}{16 \nu} \left(\frac{r_{\initial}}{M_{\MBH}}\right)^{5/2}
\left\{1-\left[1 - \frac{256}{5} \nu \left(\frac{T_{\obs} - T_{\float,\final}}{M_{\MBH}}\right) \left(\frac{M_{\MBH}}{r_{\initial}}\right)^{4} \right]^{5/8} \right\}
\nn \\
&+ \frac{5}{168 \nu^{2}} \left(\frac{r_{\final}}{M_{\MBH}}\right)^{7/2}
\left\{\left[1 + 96 \nu^{2} \left(\frac{T_{\float,\final}}{M_{\MBH}}\right) \left(\frac{M_{\MBH}}{r_{\final}}\right)^{5} \right]^{7/10} -1 \right\}\,.
\label{GW-floating-final}
\end{align}
\end{widetext}

\begin{figure}
\epsfig{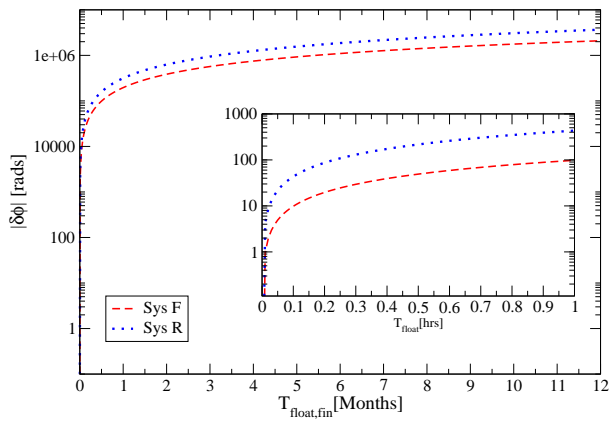}
 \caption{\label{fig:floating-dephasing-new} Dephasing in radians for Systems F (top) and R (bottom) as a function of the floating time $T_{\float,\final}$, which can be mapped to $\mu_{s}$ via Eq.~\eqref{Tfloat}.}
\end{figure}

The dephasing is then nothing but the difference between Eq.~\eqref{GW-floating-final} and Eq.~\eqref{GR-GW-phase-for-EMRIs-2} with $T_{\final,\initial} = T_{\obs}$. Such a gravitational wave dephasing depends on the usual EMRI parameters and also on the floating time $T_{\float}$. Notice that as the floating time goes to zero, Eq.~\eqref{GW-floating-final} reduces exactly the GR gravitational wave phase evolution, and thus, the dephasing vanishes. 
Figure~\ref{fig:floating-dephasing-new} plots this dephasing as a function of the floating time $T_{\float}$ for  Systems F and R. Observe how steep the dephasing is with the floating time, reaching values larger than $10^{2}$ rads in a single hour of floating evolution. 

Such large dephasings would force a template-based, matched-filtering search for EMRIs that uses GR templates to simply miss the signal all together. The only way this dephasing could become small is if $\Omega_{\float}$ is very close or higher than the highest EMRI frequency observed: 
\be
f^{\GW}_{{\rm high}} \sim \max\left[f_{\ISCO}^{\GW},f_{\rm cut,high} \right]\,,
\ee
where $f_{\rm cut,high}$ is the maximum frequency resolvable by the instrument. If $\Omega_{\float} < \pi f^{\GW}_{{\rm high}}$, then the floating resonance would prevent detection of the EMRI in the first place. Notice that if $\Omega_{\float} \lesssim \pi f^{\GW}_{{\rm low}}$, where $f^{\GW}_{\rm low} \sim 10^{-4} \; {\rm{Hz}}$ is LISA's low frequency cut-off, then the floating would last a time comparable to the Hubble time, and thus, the EMRI would never enter the LISA band in the first place. 

Given an EMRI observation that is consistent with GR, we can then automatically place a constraint on the size of $\mu_{s}$. This is because if a detection is consistent with GR, then the floating resonance had to occur at a frequency which is extremely close, or higher than, $f^{\GW}_{{\rm high}}$. Since $\Omega_{\float} \sim \Omega_{\res} \sim \mu_{s}$, this implies the constraint $\mu_{s} \gtrsim \pi f^{\GW}_{{\rm high}}$ and $\alpha$ constrained by the floating orbit condition of Eq.~\eqref{alpha_crit}. 

These results predict strong projected constraints on the $(\mu_{s},\alpha)$ phase space shown in Fig.~\ref{fig:floating-regions}. For example, the detection of a System R EMRI would rule out the entire region delimited by the curve ABC in Fig.~\ref{fig:floating-regions} (and extending further above and on the left of the plot axis). This is an improvement over Solar System constraints of over ten orders of magnitude.

\section{Conclusion and Discussion}
\label{sec:conclusions}

We have investigated whether scalar-tensor theories can be constrained with future EMRI gravitational wave observations. The gravitational and scalar emission from a EMRI system in a generic (albeit analytic) scalar-tensor theory ultimately depends on only two parameters, $(\alpha,\mu_s)$. Brans-Dicke theory is recovered in a particular case when $\mu_s=0$. In such theories, scalar dipole emission can activate and modify the radiation-reaction force via corrections to the energy flux carried away from the binary. 

We first studied massless scalar tensor theories ($\mu_s=0$ case) and numerically calculated the energy flux carried to spatial infinity and into the supermassive black hole horizon by the scalar field perturbation. We obtained this flux in the Teukolsky formalism, by solving certain master equations for a set of given quasi-circular, equatorial geodesics. We then extended the $2.5$PN expansion of this flux to $3.5$PN order through fitting coefficients. We also calculated the Fourier transform of the response function in the stationary-phase approximation beyond Newtonian order. We found that the Brans-Dicke corrected Fourier phase in the PN expansion can easily be mapped to the ppE framework of~\cite{Yunes:2009ke,2011PhRvD..84f2003C} and we derived the mapping.

Once the massless scalar energy flux had been calculated, we used it in a modified EOB-EMRI framework to estimate the dephasing induced by the massless scalar-tensor terms. We found that EMRIs do not lead to larger scalar-tensor corrections, but instead, the projected bounds on $\alpha$ (or $\omega_{\BD}$) are slightly worse than those derived from comparable-mass, early inspiral studies~\citep{Will:1994fb,Will:2004xi,Berti:2005qd,Stavridis:2010zz,Arun:2009pq,Keppel:2010qu,Yagi:2009zm}. In EMRIs, the small companion spends all of the observation time very close to the supermassive black hole. Scalar-tensor theories, however, are not necessarily strong-field modifications to GR, but rather weak-field ones. That is, the main scalar-tensor modification to the orbit is through the introduction of pre-Newtonian corrections to the radiation-reaction force due to scalar dipole emission, which is dominant for orbits with large separations.

We then studied massive scalar tensor theories ($\mu_s\neq0$ case) and confirmed that certain resonances in the scalar horizon flux arise, leading to floating orbits, and thus, drastic modifications in the gravitational wave phase evolution. During floating, the small compact object continues to inspiral, but at a rate much slower than the standard GR one. The floating inspiral is driven by the shedding of mass and spin angular momentum of the supermassive black hole. The timescale for floating to end depends on the frequency at which floating occurs, but it can easily be greater than $10^{2}$ years. In fact, if the EMRI separation is greater than $\sim 326 M_{\MBH}$, this timescale exceeds the Hubble time. 

Such large time scales suggest an easy constraint on the parameters of massive scalar-tensor theories. That is, if these resonances existed, then we should have found many systems orbiting at resonant radii, where material that would otherwise be swallowed by the supermassive black hole would accumulate. The prototypical system that comes to mind are accretion disks, where a supermassive black hole accretes material that is driven inwards by viscosity (scalar dipole emission is suppressed in neutron star/pulsar binaries due to similar sensitivities). Focusing on the orbits of accretion disk material in the absence of an EMRI, scalar flux resonances could balance the viscous force that the disk would otherwise experience, leading to a stagnation of orbits and a build up of material at a given radius (see eg.~\cite{Yunes:2011ws,Kocsis:2011dr}). Such a build up would translate into a sharp feature in the photon energy flux observed, a feature that in fact is not currently needed to fit observations~\cite{Li:2004aq,Shafee:2005ef,Steiner:2009af,McClintock:2011zq,Kulkarni:2011cy}. 

One could argue that this already rules out a certain area of parameter space of massive scalar-tensor theories, but it is not clear whether such tests can be cleanly applied. First, a continuous ring-like, configuration of particles in a circular orbit does not lead to superradiance, as only the $m=0$ modes are excited. More work is necessary to understand what happens to high $(l,m)$ modes that could lead to superradiance when the ring-like configuration is not continuous, with gaps of the size of the orbiting particles. Second, the current observational fits of the accretion disk energy spectrum are somewhat susceptible to uncertainties in accretion disk physics, requiring the inclusion of non-thermal features~\cite{Steiner:2009af} and deviations from the Novikov-Thorne disk model~\cite{Kulkarni:2011cy}. Therefore, it is not clear whether accretion disks would feel such floating resonances, and if so, whether accretion disk observations could help us constrain massive scalar-tensor theories in a clean manner. 

EMRIs, on the other hand, are much cleaner systems, and thus, their observation could be used to place constraints on massive scalar-tensor theories. For a resonance to affect an EMRI gravitational wave, the former must occur at a radius small enough that an orbiting EMRI can cross it. In the range of scalar masses in which this occurs, the floating orbit introduces an enormous dephasing in the gravitational waveform relative to the GR expectation. Such a large dephasing would prevent the detection of such modified EMRI waves with LISA or any other future detector if one match-filters with GR templates. Therefore, given a future EMRI detection that is consistent with GR, one could rule out a large area of scalar-tensor parameter space. Such projected constraints are several orders of magnitude more stringent than current Solar System ones.

Floating EMRIs emit almost monochromatic gravitational waves, and thus, one might worry that such waves could be detected as a continuous source. These waves are almost monochromatic because the radiation-reaction force becomes second-order in the mass ratio during floating, and since this quantity is much smaller than unity, the inspiral evolution is greatly slowed down. Continuous algorithms might detect such monochromatic waves, but their frequency is much higher than that of white-dwarf binaries, the traditional LISA continuous sources. In fact, there are no continuous, monochromatic sources that would emit in the high-frequency end of the LISA band according to GR. Thus, a continuous source detection at a high gravitational wave frequency could indicate the presence of floating EMRIs. 

Future work could concentrate on extending the results of this paper to relax the condition $\mu_{s} M_{\MBH} \ll 1$. Floating orbits can occur near the ISCO of a spinning black hole if $\mu_s M_\MBH\sim0.5$. Our results strongly suggest that, also in this case, floating orbits would lead to a very large dephasing of the gravitational waveform. However, a more detailed quantitative analysis, possibly involving the numerical fluxes we have here presented, would be desirable. One could also investigate whether the non-floating resonances have structure and how this modifies the waveform observable if the resonant width were large enough. 

Another possible avenue for future research would be to investigate whether additional constraints can be placed on scalar-tensor theories given other astrophysical observations. For example, if floating orbits exist, then the spin of the supermassive black hole would be decreased after every EMRI coalescence. Statistically speaking, one would then predict a supermassive black hole spin distribution that is peaked close to zero. Therefore, if a sufficient number of supermassive black holes are observed with sufficiently large spins, one would then infer that if floating orbits exist, then they had to be hidden inside the horizon of the supermassive black holes. This can then be used to place a statistical constraint on the coupling parameters of scalar-tensor theories~\cite{Bence}. 

Other future work could focus on determining whether the floating resonances found here are generic to other theories. For example, recently there has been much effort to understand certain quadratic gravity theories~\cite{Yagi:2011xp}, where the Einstein-Hilbert action is enhanced through the addition of terms composed of the product of a scalar field and all quadratic curvature invariants. Usually, the scalar field in these theories is assumed massless, but if it were endowed with a mass, superradiant-induced, floating resonances could also arise. If so, the same mechanism described here to constrain the mass of scalar-tensor theories could be used to bound the mass of quadratic-gravity scalar fields.  

\acknowledgments

We would like to thank Emanuele Berti, Cliff Burgess, Michael Horbatsch, Scott Hughes, Bence Kocsis, Avi Loeb, Saeed Mirshekari, Takahiro Tanaka, Clifford Will and Kent Yagi for useful comments and suggestions. NY acknowledges support from NSF grant PHY-1114374 and NASA grant NNX11AI49G, under sub-award 00001944. NY also acknowledges support from NASA through the Einstein Postdoctoral Fellowship Award Number PF0-110080, issued by the Chandra X-ray Observatory Center, which is operated by the Smithsonian Astrophysical Observatory for and on behalf of NASA under contract NAS8-03060. PP and VC acknowledge support from the {\it DyBHo--256667} ERC Starting Grant, and from FCT - Portugal through PTDC projects FIS/098025/2008, FIS/098032/2008, CTE-AST/098034/2008, CERN/FP/123593/2011. PP acknowledges financial support provided by the European Community through the Intra-European Marie Curie contract aStronGR-2011-298297. The authors also thankfully acknowledge the computer resources, technical expertise and assistance provided by the Barcelona Supercomputing Centre---Centro Nacional de Supercomputaci\'on, as well as by the DEISA Extreme Computing initiative and the Milipeia cluster in Coimbra. This work was granted access to the HPC resources of PSNC in Poland made available within the Distributed European Computing Initiative by the PRACE-2IP, receiving funding from the European Community's Seventh Framework Programme (FP7/2007-2013) under grant agreement nr RI-283493.

\appendix
\section{Flux resonances in the low-frequency, low-mass regime}
\label{app:resonances}
In this appendix we derive Eq.~\eqref{res-height} and we generalize it for any $(l,m)$ mode.
For simplicity, we set $M_\MBH=1$ and all our intermediate derivations are done using dimensionless quantities. At the end of the calculation, our results have a simple scaling with $M_\MBH$ and it is easy to re-insert the necessary factors, as we do in Eq.~\eqref{res-height_gen}.

The procedure is based on solving Eq.~\eqref{nonhom0} by the method of matched asymptotic expansions~\cite{Bender,Detweiler:1980uk}. At large distance, the wave equation~\eqref{nonhom0} is approximately
\be
\frac{d^2 X_{lm\omega}}{dr^2}+\left(k_{\infty}^2+\frac{2\mu_s^2}{r}-\frac{l(l+1)}{r^2}\right)X_{lm\omega}=0\,.\nn
\ee
The solution of the equation above with the correct behavior at infinity reads
\be
X_{lm\omega}^{\infty}=r^{l+1}e^{ik_{\infty}r}U\left(l+1- i \frac{\mu_s^2}{k_{\infty}},2l+2,-2ik_{\infty}r\right)\,,\nn
\ee
where $U(a,b;z)$ is the hypergeometric function.
This solution can be expanded for small distances as
\begin{eqnarray}
X_{lm\omega}^{\infty}&&\sim \frac{\Gamma[-1-2l]}{\Gamma[-l-i\mu_s^2/k_{\infty}]}r^{l+1}+\nn\\
&&-(-2ik_{\infty})^{-2l-1}\frac{\Gamma[1+2l]}{\Gamma[l+1-i\mu_s^2/k_{\infty}]}r^{-l}\,.\label{wavefunction1}
\end{eqnarray}
Following Detweiler~\cite{Detweiler:1980uk}, we define $i\mu_s^2/k_{\infty}=l+1+n+\delta \nu$. Close to the resonances, $ik_{\infty}\sim -\mu_s^2/(l+1+n)$ and $\delta \nu\sim 0$. Therefore $U(l+1-\mu_s^2/k_{\infty},2l+2,-2ik_{\infty}r)\sim 1$ and the wavefunction at the resonance reads
\be
X^{\infty}_{lm\omega}\sim r_{\res}^{l+1}e^{-\mu_s^2 r_{\res}/(l+1+n)}\,.\label{behaviorresonance}
\ee
We have verified this numerically and we find remarkable agreement between Eq.~\eqref{behaviorresonance} and the waveform obtained by integrating numerically Eq.~\eqref{nonhom0}, all throughout the domain.

On the other hand, close to the horizon, we can approximate the wave equation as
\be
z(z+1)\frac{d}{dz}\left[z(z+1)\frac{dR}{dz}\right]+\left[{\cal K}^2-l(l+1)z(z+1)\right]R=0\,,\nn
\ee
where $X_{lm\omega}=\sqrt{r^2+a_\MBH^2} R(z) $, $z\equiv(r-r_+)/(r_+-r_-)$ and ${\cal K}$ is defined below Eq.~\eqref{res-height}.
The solution to the above equation reads
\be
R=(-1)^{i{\cal K}}\left (A \, P_l(2i{\cal K},1+2z)+B \, Q_l(2i{\cal K},1+2z)\right)\,,\nn
\ee
where $P_l$ and $Q_l$ are Legendre polynomials and Legendre functions of second kind, respectively. At large distances, the near-region wave-function reads
\be
X_{lm\omega}\sim q_1r^{-l}\left(A+\frac{Bi\pi}{2}\frac{3+e^{4\pi {\cal K}}}{1-e^{4\pi {\cal K}}}\right)+q_2r^{l+1}\left(A-\frac{Bi\pi}{2}\right)\,,\label{wavefunction2}
\ee
with
\begin{eqnarray}
q_1&=&\frac{(-1)^{l+1}}{4(l+1/2)}\frac{\Gamma[l+1]\,(r_+-r_-)^{l+1}}{\Gamma[2l+1]\Gamma[-l-2i{\cal K}]}\,,\\
q_2&=&\frac{\Gamma[1+2l]\,(r_+-r_-)^{-l}}{\Gamma[l+1]\Gamma[l+1-2i{\cal K}]}\,.
\end{eqnarray}
%
Thus, by matching the two wavefunctions~\eqref{wavefunction1} and \eqref{wavefunction2} we can extract the coefficients $A,B$, which read
\begin{widetext}
\begin{eqnarray}
A&=&\frac{(-1)^{-l} \sqrt{\pi}(-i k_\infty)^{-2 l} }{4 \Gamma\left[\frac{1}{2}+l\right](r_+-r_-)^{l}} \left[\frac{\left(3+e^{4 {\cal K} \pi }\right)  \Gamma[-1-2 l] \Gamma[1+l-2 i {\cal K}]}{4^{l} \left[k_\infty(r_+-r_-)\right]^{-2 l} \Gamma\left[-l-\frac{i  \mu_s^2}{k_\infty}\right]}+\frac{\left(1-e^{4 {\cal K} \pi }\right) \Gamma\left[\frac{1}{2}+l\right]^2 \Gamma[2+2 l] \Gamma[-l-2 i {\cal K}]}{ik_\infty \pi(r_+-r_-) \Gamma\left[1+l-\frac{i  \mu_s^2}{k_\infty}\right]}\right]\,,
\label{coeffA} \\
B&=& \frac{\left(-1+e^{4 {\cal K} \pi }\right) k_\infty^{-2 l} }{2 \pi ^{3/2} \Gamma\left[\frac{1}{2}+l\right](r_+-r_-)^{l}}\left[\frac{\Gamma\left[\frac{1}{2}+l\right]^2 \Gamma[2+2 l] \Gamma[-l-2 i {\cal K}]}{k_\infty (r_+-r_-) \Gamma\left[1+l-\frac{i  \mu_s^2}{k_\infty}\right]}-\frac{i  \pi k_\infty^{2l}\Gamma[-1-2 l] \Gamma[1+l-2 i {\cal K}]}{4^{l} (r_+-r_-)^{-2 l} \Gamma\left[-l-\frac{i  \mu_s^2}{k_\infty}\right]}\right] \,. \label{coeffB}
\end{eqnarray}
\end{widetext}

Now that the solutions have been asymptotically matched, let us compute the Wronskian. 
Since the Wronskian of two independent solutions, as defined below Eq.~\eqref{Z}, is constant throughout the space, we can evaluate it at an arbitrary point. In particular, in the near-horizon region it reads 
\be
W = 2ik_+B_{\rm out}\,,\label{Wdef2}
\ee
where we have used the definition below Eq.~\eqref{Z} and the asymptotic behavior of the homogeneous solutions close to the horizon,
\begin{eqnarray}
 X_{lm\omega}^{r_+}&\sim& e^{-i k_+ r_*} \,\\
 X_{lm\omega}^{\infty}&\sim& A_{\rm in} e^{-i k_+ r_*} + B_{\rm out} e^{i k_+ r_*} \,. \label{Xhomo_nearhor}
\end{eqnarray}
Hence, the Wronskian is simply related to the coefficient of the outgoing wave of $X_{lm\omega}^{\infty}$ in the near-horizon region.
Using Eq.~\eqref{wavefunction2} and the coefficients in Eq.~\eqref{coeffA} and~\eqref{coeffB},the solution $X_{lm\omega}=X_{lm\omega}^{\infty}$ close to the horizon becomes
\ba
X_{lm\omega}^{\infty}&\sim&\frac{\sqrt{r_+^2+a_\MBH^2}}{\Gamma[1-2i{\cal K}]}z^{-i{\cal K}}\left(A+\frac{i \pi B}{2}\frac{1+e^{4\pi {\cal K}}}{1-e^{4\pi{\cal K}}}\right)
\nonumber \\
&+&\sqrt{r_+^2+a_\MBH^2}z^{i{\cal K}}
\nonumber 
\\ 
\nonumber  
&\times&
\frac{i B\pi^2 \left[(1-\tanh^{-1}[2 \pi{\cal K}]) \sec^{-1}[l \pi +2 i\pi{\cal K} ]\right]}{2\Gamma[-l-2i{\cal K}]\Gamma[1+l-2i{\cal K}]\Gamma[1+2i{\cal K}]}\,.\\ \label{Xnearhor}
\ea
The first term in this expansion is in fact an outgoing wave because, close to the horizon,
\be
(r-r_+)^{\pm\frac{(r_+^2+a_\MBH^2)ik_+}{r_+-r_-}}=(r-r_+)^{\mp i{\cal K}}=e^{\pm i k_+r_*}\,,\nn
\ee
so that $z^{-i{\cal K}}=e^{ik_+r_*}(r_+-r_-)^{i{\cal K}}$. Therefore, Eq.~\eqref{Xnearhor} has exactly the same form of Eq.~\eqref{Xhomo_nearhor} and one can easily extract the coefficient $B_{\rm out}$. With this at hand, the Wronskian~\eqref{Wdef2} is simply
\begin{eqnarray}
W&=&2ik_+B_{\rm out}=2ik_+\frac{\sqrt{r_+^2+a_\MBH^2}(r_+-r_-)^{i{\cal K}}}{\Gamma[1-2i{\cal K}]}\times\nn\\
&\times&\left(A+\frac{i\pi B}{2} \frac{1+e^{4\pi {\cal K}}}{1-e^{4{\cal K}\pi}}\right)\,.\nn
\end{eqnarray}
and using Eqs.~\eqref{coeffA} and \eqref{coeffB} we find 
\begin{eqnarray}
W&=&\frac{k_+ \sqrt{a_\MBH^2+r_+^2} (r_+-r_-)^{-l+i {\cal K}}}{\sqrt{\pi } \Gamma\left[\frac{1}{2}+l\right] \Gamma[1-2 i {\cal K}]} \times\nn\\
&\times&\left(\frac{\Gamma\left[\frac{1}{2}+l\right]^2 \Gamma[2+2 l] \Gamma[-l-2 i {\cal K}]}{k_\infty^{1+2 l} (r_+-r_-) \Gamma[-n-\delta \nu ]}+\right.\nonumber\\
&&\left.+\frac{i  \pi  (r_+-r_-)^{2 l} \Gamma[-1-2 l] \Gamma[1+l-2 i {\cal K}]}{4^{l}\Gamma[-1-2 l-n-\delta \nu ]}\right)\,,\nn
\end{eqnarray}
which is valid for any $(l,m,n)$. Focusing on the fundamental mode, $n=0$ only, at resonance $\delta\nu\to0$ and $W$ takes the form 
\be
W_\res=\frac{ik_+ \sqrt{r_+^2+a^2_\MBH} \Gamma[l+1]\Gamma[1+l-2 i {\cal K}]}{(r_+-r_-)^{-l-i {\cal K}}\Gamma[2l+1]\Gamma[1-2 i {\cal K}]}\label{Wres}\,.
\ee
For $l=1=m$, this simply reduces to
\be
W_\res=k_+\left({\cal K}+\frac{i}{2}\right)\sqrt{r_+^2+a^2_\MBH}(r_+-r_-)^{1+i{\cal K}}\,.
\ee

Finally, we can now estimate the peak flux coming out of the horizon at the resonant frequencies and in the small $\mu_s$ limit. Using Eqs.~\eqref{scalar_fluxes} and \eqref{Z}, it reads
\be
\dot E_\text{peak}^{s}\sim \frac{\alpha^2 \nu^2 \omega_\res k_+\,|S_{l'm'}^{*}(\pi/2)|^2}{r_\res^2}\frac{\left|X_{lm\omega}^{\infty}\right|^2}{|W_\res|^2}
\ee
Using (\ref{behaviorresonance}) we find
\be
\dot E_\text{peak}^{s}\sim \frac{\alpha^2 \nu^2 \omega_\res k_+\,|S_{l'm'}^{*}(\pi/2)|^2}{r_\res^2}\frac{r_\res^{2(l+1)}e^{-2\mu_s^2  r_\res/(l+1)}}{|W_\res|^2}
\ee
On the other hand at large distances, from Eq.~\eqref{om_resonance}, $\omega_\res^2 \sim r_\res^{-3}\sim \mu_s^2$, and the expression above simplifies to
\be
\dot E_\text{peak}^{s}\sim {\alpha^2 \nu^2 k_+\,|S_{l'm'}^{*}(\pi/2)|^2}\frac{r_\res^{2l-3/2}e^{-2\mu_s^2  r_\res/(l+1)}}{|W_\res|^2}\label{flux_ana_fin}\,,
\ee
where $|W_\res|^2$ is given by Eq.~\eqref{Wres}. The final expression, restoring the necessary factors of $M_\MBH$, reads
\begin{widetext}
 \begin{equation}
\dot E_\text{peak}^{s}\sim  \frac{\alpha^2 \nu^2\, m \, {\cal K} \,\pi|S_{lm}(\pi/2)|^2 \Gamma[1+2 l]^2  \left(1-q_\MBH^2\right)^{-l}}{4^{l}  \Gamma[1+l]^2 \Gamma[1+l-2 i {\cal K}] \Gamma[1+l+2 i {\cal K}]k_+ r_+}\frac{r_\res^2/M_\MBH^2+2 q_\MBH \sqrt{ r_\res/M_\MBH}-3 r_\res/M_\MBH  }{ \left(q_\MBH+(r_\res/M_\MBH)^{3/2}\right)^3 \left(q_\MBH^2+r_\res^2/M_\MBH^2\right) }\left(\frac{r_\res}{M_\MBH}\right)^{3+2 l} \,.\label{res-height_gen}
\end{equation}
\end{widetext}
We have checked this expression against numerical solution and found very good agreement in all of parameter space for $\mu_s M_\MBH\ll1$. 
For $l=m=1$, Eq.~\eqref{res-height_gen} reduces to Eq.~\eqref{res-height} in the main text.

\bibliography{master}
\end{document}